\documentclass{article}

\usepackage{listings,xcolor}

\definecolor{codegreen}{rgb}{0,0.6,0}
\definecolor{codegray}{rgb}{0.5,0.5,0.5}
\definecolor{codepurple}{rgb}{0.58,0,0.82}
\definecolor{backcolour}{rgb}{0.95,0.95,0.92}

\lstdefinestyle{mystyle}{
    backgroundcolor=\color{backcolour},   
    commentstyle=\color{codegreen},
    keywordstyle=\color{magenta},
    numberstyle=\tiny\color{codegray},
    stringstyle=\color{codepurple},
    basicstyle=\ttfamily\footnotesize,
    breakatwhitespace=false,         
    breaklines=true,                 
    captionpos=b,                    
    keepspaces=true,                 
    numbers=left,                    
    numbersep=5pt,                  
    showspaces=false,                
    showstringspaces=false,
    showtabs=false,                  
    tabsize=4
}
\usepackage{graphicx} % Required for inserting images
\usepackage{amsmath,amssymb,epsf} \usepackage{latexsym,bm,
  slashed} 
  \usepackage{xcolor} 
  
  \usepackage{authblk}
  \definecolor{dark}{rgb}{0.10,0.2,0.3}
\definecolor{magenta}{rgb}{0.7,0.1,0.3}
\definecolor{purpure}{rgb}{0.5,0.15,0.3}
\usepackage[font=small,format=plain,labelfont=bf,up,textfont=it,up]{caption}
\usepackage{hyperref, cite} 
\hypersetup{colorlinks, citecolor=blue,
  filecolor=blue, linkcolor=magenta,
  urlcolor=purpure,hyperfootnotes=true,pdftex} 
  
\newcommand{\tr}{{\rm tr}}

\newcommand{\rt}{{\bm r}}

\newcommand{\xt}{\mathbf{x}}
\newcommand{\asb}{\bar{\alpha}_s}

\usepackage{amssymb}

\title{Deep inelastic scattering as a probe of entanglement:\\ the complete QCD dipole cascade}

\author[1]{Martin~Hentschinski}
\author[2]{Krzysztof~Kutak}
\author[3]{Wies{\l}aw P{\l}aczek}
\author[4]{Martin Rohrmoser}

\affil[1]{\normalsize Departamento de Actuaria, F\'isica y Matem\'aticas, 
Universidad de las Americas Puebla, San Andr\'es Cholula, 72820 Puebla, Mexico }
\affil[2]{ \normalsize Institute of Nuclear Physics, Polish Academy of Sciences, 
 ul.~Radzikowskiego 152, 31-342 Krak\'ow, Poland}
 
 \affil[3]{ \normalsize Faculty of Physics Astronomy and Applied Computer Science and Mark Kac Center for Complex Systems Research, Jagiellonian University, ul.\ {\L}ojasiewicza 11,
30-348 Krak\'ow, Poland}
\affil[4]{ \normalsize Institute of Physics, Jan Kochanowski University, ul.\ Uniwersytecka 7, 25-406 Kielce, Poland}

\begin{document}

\maketitle
\begin{abstract}
 We study entanglement entropy in Deep Inelastic Scattering (DIS) using the dipole formulation of the high-energy limit of QCD. We argue that a reduced density matrix arises in low $x$ DIS  due to a trace over unobserved color degrees of freedom and we obtain entanglement entropy in terms of dipole multiplicities, directly from the von Neumann entropy.
 %In the present study we explicitly construct a density matrix representing the entangled state of the photon and proton and perform a trace over unobserved degrees of freedom.
Dipole multiplicities are  obtained from a solution to  low $x$ evolution equations, which  we solve numerically. Unlike previous studies, we take into account  both transverse-size and azimuthal-angle dependence in the dipole evolution kernel. We study both the exact solution of the equation as well as its double leading-logarithmic approximation (DLLA). We find that for the same initial dipole size, the DLLA solution generates a larger entropy.
%We interpret this as information loss due to erasing data on daughter dipole sizes.
Finally, we calculate the dipole multiplicities and entanglement entropy and compare our results to  the Shannon entropy of hadron multiplicities, as measured  by the H1 collaboration.
\end{abstract}

\section{Introduction}

In recent years, there has been growing interest in studies of entanglement in association with collider physics \cite{Bauer:2022hpo,Abir:2023fpo,Afik:2025ejh}. These studies address entanglement in top-quark and Higgs-boson decays, spin correlations in quark-antiquark production, and the relation between spin and angular momentum \cite{Afik:2025ejh,DelGratta:2025qyp,STAR:2025njp,Maltoni:2024csn,Aoude:2023hxv,Datta:2024hpn,Florio:2025hoc,Qi:2025onf,Hatta:2024lbw,Altomonte:2024upf,Gu:2025ijz,Fucilla:2025kit,Fucilla:2026mkg,Amorosso:2026mdo,Liu:2026ees,Liu:2026dzv,Gautam:2026rek}. In these studies, various quantum measures such as concurrence, magic, and entropy are used to determine to what extent the final-state products are entangled. In the present paper, we continue the analysis of the proposal \cite{Kharzeev:2017qzs} to study entanglement in deep inelastic scattering (DIS) of electrons on protons. 
With $t= - Q^2$ the moment transferred between electron and proton and $1/Q$ significantly smaller than the extension of the proton, the virtual photon probes the partonic content of the proton and effectively performs a partial measurement of the proton wave function. 
%where it has been argued that in DIS, 
%where the virtual photon probes the partonic content of the proton, one effectively performs a partial measurement of the proton wave function. 
%Incomplete information
With the proton itself a highly entangled quantum state, partial observation of the proton wave function %can be characterized by 
gives rise to entanglement entropy. 
Once produced, this entropy will also reflect itself in the hadronic final state and it has been argued in \cite{Kharzeev:2021nzh,Kharzeev:2026jkq} that entanglement entropy can be   experimentally accessed through hadronic entropy of the  final state of collision, which is obtained from the measured multiplicity distributions of charged hadrons.  First phenomenological evidence for this proposal has been provided in \cite{Hentschinski:2022evidence,Hentschinski:2023maxent,Hentschinski:2024qcd_evo,Hentschinski:2025pyq}.
%which 
%is directly related to hadronic entropy
%in the final state of the collision can be obtained as the hadron entropy calculated from measured multiplicity distributions of charged hadrons.
For related works addressing entropy production in high energy physics, see \cite{Kutak:2011rb,Peschanski:2012cw,Stoffers:2012mn,Kovner:2015hga,Kovner:2018rbf,Peschanski:2019yah,Hagiwara:2017uaz,Chachamis:2023omp,Berges:2017hne,Kutak:2025tsx,Golec-Biernat:2025hwa,Ouchen:2025ooo, Rabelo-Soares:2025ams, Lokos:2025cbu,Ouchen:2026blf,Ouchen:2026ymw,Kharzeev:2026inq}.

The proposal \cite{Kharzeev:2017qzs} was presented using the degrees of freedom of the high-energy limit of QCD. In the high-energy or Regge limit of QCD, i.e.\ where the collision energy is the largest scale as compared to masses and the involved momenta, gluons and sea quarks are the dominant parton degrees of freedom. Provided that the occupation numbers of quarks and gluons are not too large, the relevant evolution equation is in this limit the Balitsky--Fadin--Kuraev--Lipatov (BFKL) equation \cite{Fadin:1975cb, Lipatov:1976zz, Kuraev:1976ge, Balitsky:1978ic, Lipatov:1985uk}. If formulated in coordinate space \cite{Mueller:1993rr,Mueller1995UnitarityBFKL}, it can be expressed as an evolution equation for gluon multiplicities, which in the large-number-of-color approximation is equivalent to an evolution equation in the number of color dipoles. 

Previous phenomenological studies address inclusive DIS within the so-called `1+0' reduction of the Mueller dipole cascade in \cite{Hentschinski:2022evidence,Hentschinski:2023maxent,Hentschinski:2024qcd_evo}, its double-leading-log approximation (DLLA) \cite{Liu:2022bru}, and Monte Carlo simulations \cite{Hentschinski:2025pyq}, where the particular role of soft gluons in the generation of entropy has been observed. The `1+0' model has been further applied to a description of entanglement entropy in Diffractive DIS \cite{Hentschinski:2023izh} as well as in proton--proton ($pp$) collisions \cite{Tu:2019ouv}. There are various extensions of the original proposal, i.e.\ to study the entropy of dipoles that split and recombine \cite{Hagiwara:2017uaz,Caputa:2024xkp,Kutak:2025syp}, allowing them to transition to the vacuum as well as to account for modification of the dipole equation that changes the underlying geometric distribution to the negative binomial distribution \cite{Caputa:2024xkp,Kutak:2025tsx}. An extension which takes into  AGK cutting rules has been presented in \cite{Ouchen:2025tta} and used to describe multiplicities in $pp$ processes. 

In the present paper, we determine entanglement entropy using the  complete Mueller dipole cascade, as originally proposed in \cite{Kharzeev:2017qzs}. The paper is organized as follows. In Section~2, we review in more detail the mechanism of entropy production in DIS and present the equations to be studied numerically, i.e.\ the BFKL equation for dipole multiplicities, its double-leading-log approximation, as well as its 1+0 reduction. We furthermore derive an equation for the mean multiplicity within the DLLA. 
In Section~3, we present the results for dipole multiplicity, its mean value and entropy as obtained from the considered equations, based on specifically developed Monte Carlo algorithms. In Section~4, we present results for the description of entropy data obtained in DIS as measured by the H1 Collaboration \cite{H1:2020zpd}. 
Section~5 contains a summary and conclusions of our work.
Appendix~A is devoted to derivation of a non-splitting term in DLLA. In Appendix~B, we describe Monte Carlo algorithms and their implementations that have been developed to solve the above equations.

\section{Evolution equations for the dipole number distribution}
\label{sec:eveq}

In this section we briefly recapitulate the underlying picture which gives rise to finite entropy in the DIS reaction, see also the original publication \cite{Kharzeev:2017qzs} on this subject, as well as \cite{Kharzeev:2021nzh, Kharzeev:2026jkq},  see also \cite{Liu:2022hto} for a related discussion.

\subsection{Decoherence and entanglement entropy in DIS at low $x$}
\label{sec:decoherence}

In the DIS reaction, the proton, with four momentum $p$, is observed through the interaction of its constituents with a highly virtual photon with four momentum $q$ and virtuality $Q^2 = -q^2 > 0$.  During this interaction, the observing system, i.e.\ the virtual photon, becomes entangled with the proton. Coherence of the proton, an initially pure quantum state\footnote{In a collider experiment the proton is accelerated through interaction with electromagnetic fields. It is assumed that the accelerated proton is still (at least to a good approximation) an eigenstate of the QCD Hamiltonian.}, is therefore lost. The resulting hadronic system constitutes a mixed quantum system and is characterized by finite entropy.  Depending on the precise form of the interaction Hamiltonian, observation of a quantum system (the proton) through another quantum system (the virtual photon) implies the emergence of a set of preferred states, so called {\it pointer states }\cite{Schlosshauer:2019ewh},  which are eigenstates of the interaction Hamiltonian.  During the interaction with the observing system, these states are selected as the natural basis to which the system will evolve to, i.e.\ in this basis, off-diagonal elements of the density matrix will vanish rapidly as a consequence of the observation. 

In the low $x$ limit of DIS, where $x = Q^2/2 p \cdot q$, a natural candidate for  pointer states is provided by color dipoles. Originally proposed  in \cite{Mueller:1993rr} for the case of heavy onia (which ensure the applicability of QCD perturbation theory), color dipoles have been identified as natural building blocks of the wave function of  a 'fast' hadron,  which can extend over several units in rapidity, up to $Y = \ln 1/x$. For the onium state, they arise within a  leading logarithmic approximation and making use of a large $N_c$ approximation, which allows to rearrange a system of a quark anti-quark pair and $n$ gluons into $n+1$ color dipoles; each color dipole is then characterized by the transverse position of its constituents (an effective quark--anti-quark system) and a pair of color indices in the (anti-)fundamental representation of SU$(N_c)$.  
%; within a  leading logarithmic approximation and making use of a large $N_c$ approximation, it is possible to express the hadron wave function in a color dipole Fock basis. 
%In more detail, a system of $n+1$ color dipoles arises from a rearrangement of a system of $q\bar{q}$ pair and $n$ gluons, where each color dipole is finally characterized by its transverse position of its constituents (an effective quark-anti-quark system).  
Assuming a configuration which justifies the use of QCD perturbation theory, e.g.\ creation of  a system of dipoles with sufficiently small transverse separation, such an approximation holds also for the proton\footnote{At the very least in the limit where parton multiplicities are large and within the leading logarithmic approximation which counts only terms where the smallness of the strong coupling constant $\alpha_s$ is compensated by a large logarithm $Y = \ln 1/x$.} and its wave function can be  approximated through an expansion in the dipole number Fock basis. In DIS, this system of color dipoles is then probed by another color dipole (or a system thereof) which originates from the virtual photon. 
%In the low $x$ limit, the virtual photon  splits long before the interaction with the proton into a quark-anti-quark pair which then interacts with dipoles in the proton through the exchange of virtual gluons, which is the dominant interaction in the low $x$ or high energy limit. 

\begin{figure}[t]
  \centering
  \includegraphics[width=.47\textwidth]{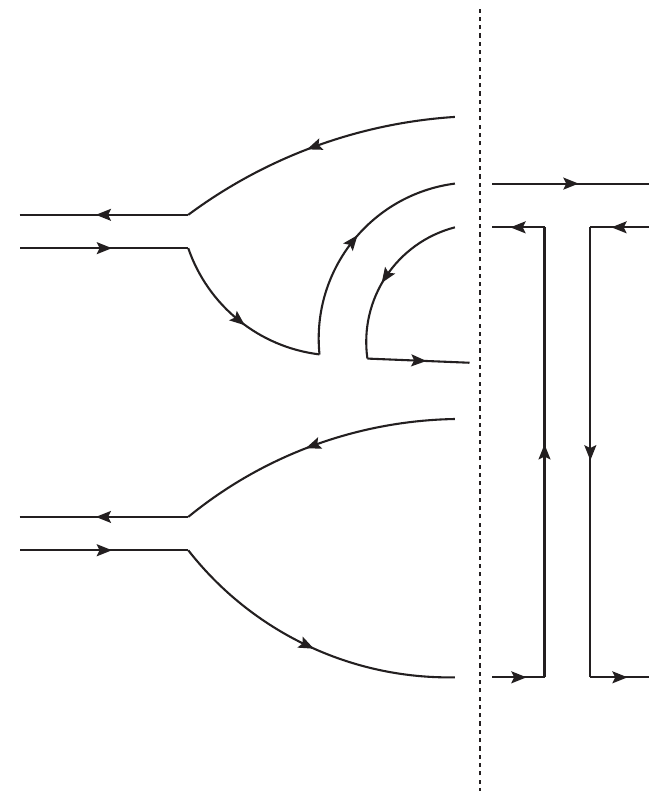}
 
  \caption{Schematic scattering of two color dipoles, e.g.\ a virtual photon and a heavy onium state; the dashed line indicates the transition from the initial-state wave function towards the color-entangled state. Each of the scattering particles maintains its transverse position during the scattering, while its color charge is being modified. Coherence of the initial wave functions is being lost. }
  \label{fig:dipole_scattering}
\end{figure}

Within high-energy factorization, the interaction Hamiltonian, which describes the interaction between both dipole systems, is such that it conserves the transverse positions of scattering partons, see e.g.\ \cite{Marchesini:2003nh}. They are  therefore a natural candidate for the pointer basis in  low-$x$ DIS, since they constitute eigenstates of the interaction Hamiltonian. Nevertheless, while the transverse position of partons in the photon and proton are conserved,  the interaction Hamiltonian is non-diagonal in color space;  each color dipole picks up a color phase, which can be expressed as a path ordered exponential in the high energy gluon field and which describes the rotation of the dipole in color space; the low-$x$ wave functions of the proton and virtual photon are therefore entangled through the low-$x$ gluon field. Decoherence of the proton and photon wave function arises then due to the modification of their color state during the interaction, see Fig.~\ref{fig:dipole_scattering}.

 The association of scattering partons with the proton or the photon wave function is of course arbitrary, in particular within the high-energy factorization and the leading logarithmic approximation. The leading order BFKL formalism only assigns a minimal number of partons to photon and proton states, e.g. a quark--antiquark pair in the case of the virtual photon, while multiple gluon production is associated with the BFKL Green's function, i.e.\ the colored system exchanged between scattering objects. A different picture is obtained if the scattering onium state is considered in the infinite momentum frame, as adapted in the original formulation of the color dipole picture \cite{Mueller:1993rr}: In this case all dipoles are contained in the onium wave function (which serves as our model of the low-$x$ proton); the virtual photon is on the other hand elementary. In this latter case, the photon virtuality $Q^2$  determines the size of the  $q\bar{q}$ dipole which interacts with the photon. Coherence of the proton wave function  is lost, since the virtual photon  triggers the manifestation of a chain of small dipoles and excludes in this way  configurations of the proton wave function which are inconsistent with such small dipole sizes. 

 For a discussion within the QCD high-energy factorization, the most natural framework is, however, scattering of two systems of color dipoles which originate from the virtual photon and the hadron, with corresponding transverse extensions. Both systems then interact through the exchange of high-energy gluons, which maintain the transverse position of scattering partons (and therefore transverse separation of dipoles), while their color state is changed\footnote{Even though the transverse position is preserved during the interaction, the interaction between systems characterized by small (virtual photon) and large (proton) dipole sizes also enhances/suppresses certain  configurations  and which gives rise to evolution in the transverse dipole size in the scattering dipole systems.}, see  Fig.~\ref{fig:dipole_scattering} for an illustration.

Independent of the frame, one therefore arrives at a state in which the photon and proton wave functions are entangled. This is of course a trivial statement, since observation of the proton through the virtual photon in the DIS reaction takes only place if both Hilbert spaces are entangled \cite{Schlosshauer:2019ewh}. Expressing now color degrees of freedom of gluons through fundamental ($i$) and anti-fundamental color indices ($j$),  the entangled DIS state of photon and hadron can be schematically expressed in the dipole Fock basis as
\begin{align}
  \label{eq:gammaH}
  |\text{DIS}\rangle & = 
\sum_{n=1}^\infty \prod_{m=1}^n \int\limits_0^1 \frac{dz_m}{z_m} \int d^2 \rt_m
\sum_{i_1\dots i_{n}} \sum_{j_1\dots j_{n}} \, 
a^{i_1\dots i_{n}}_{j_1\dots j_{n}}\left( \{z_l, \rt_l\}\right)
|\{z_l, \rt_l\}\rangle |i_1\dots i_{n} \rangle |j_1\dots j_{n}\rangle.
\end{align}
Here $\{z_l, \rt_l\} \equiv z_1, \ldots z_n, \rt_1, \ldots \rt_n$, while the sum contracts color indices in (anti-)color kets and coefficients $a$. $\rt_l$ denotes the transverse size of the $l$th dipole, while $z_l$ is the proton momentum fraction that can be associated with the $l$th dipole, see \cite{Mueller:1993rr} for details. The coefficients $a$ depend furthermore on the momenta of the virtual photon and proton. Within the leading logarithmic approximation, this dependence is largely suppressed  and replaced by corresponding regulators, which are then associated with the appropriate kinematic limits, e.g.\ $1 > z_l > x$, etc. We will specify those details below in the explicit discussion of evolution equations. 

 The density operator  is obtained as an outer product of the wave function, $\rho = | \text{DIS}\rangle  \langle \text{DIS}|$. Due to color confinement, it is not possible to observe the colored states of individual partons, and we therefore need to  average over color degrees of freedom.
Indeed, during the measurement of a certain observable $\mathcal{O}$  in the DIS reactions,  we always extract the quantum mechanical expectation value $\langle \mathcal{O}\rangle = \tr \left(\rho \mathcal{O} \right)$, since we commonly average over many scattering events. Since experiments cannot observe color degrees of freedom, the operator $\mathcal{O}$ does not act on color space; therefore, the relevant density matrix is the reduced density matrix with color degrees of freedom traced over: 
\begin{align}
  \label{eq:trace_color}
\rho_R  & = \langle  \rho\rangle_{\text{color}}  =   \tr_{\text{color}} \rho =  \sum_{n=1}^\infty \langle i_1, \ldots i_{n}| \langle j_1 \ldots j_{n} | \rho | i_1, \ldots i_{n} \rangle | j_1  \ldots j_{n}\rangle \notag \\
& = 
\sum_{n=1}^\infty \prod_{m=1}^n \int\limits_0^1 \frac{dz_m}{z_m} \int d^2 \rt_m
\prod_{k=1}^n \int\limits_0^1 \frac{dz_k'}{z_k'} \int d^2 \rt_{k}' \sum_{i_1\dots i_{n}} \sum_{j_1\dots j_{n}}  \, 
 |\{z_l, \rt_l\}\rangle  \langle \{z_{l}', \rt_{l}'\} |
 \notag \\
 & \hspace{8cm}   a^{i_1\dots i_{n}}_{j_1\dots j_{n}}\left( \{z_l, \rt_l\}\right)  a^{i_1\dots i_{n}*}_{j_1\dots j_{n}}\left( \{z_{l}', \rt_{l}'\}\right)
 .
\end{align}
Due to the average over color degrees of freedom, the  reduced density matrix is now diagonal in the number of partons. Note that this density matrix does not automatically coincide with the density matrix of the observed hadronic system. In particular, it does not include hadronization, which itself provides an average over color degrees of freedom, albeit different from the above implementation.  Nevertheless, within the high-energy limit, interaction takes place during a very short time interval and interactions between dipoles which extend over large rapidity intervals after the initial scattering are suppressed. It is therefore natural to assume that the reduced density matrix provides at the very least a good approximation to the density matrix of the hadronic system,  in particular for observables, which are invariant under unitary transformations and therefore time evolution of the entire system, such as entanglement entropy. The latter can now be determined using the replica trick. With
\begin{align}
\label{eq:trrho_R}
\tr \rho_R^r & = \sum_n p_n^r,
&  p_n & =  \prod_{m=1}^n  \int\limits_0^1 \frac{dz_m}{z_m} \int d^2 \rt_m \sum_{i_1\dots i_{n}} \sum_{j_1\dots j_{n}}  \left|a^{i_1\dots i_{n}}_{j_1\dots j_{n}}\left( \{z_l, \rt_l\}\right) \right|^2,
\end{align}
and one finds for the entanglement entropy
\begin{align}
  \label{eq:replica}
  S & = \lim_{r \to 1} \frac{1}{r-1} \ln \tr \rho_R^r =- \lim_{\epsilon \to 0} \frac{1}{\epsilon}  \ln \tr \rho_R^{1+\epsilon} = -\sum_n p_n \ln p_n, 
\end{align}
which corresponds to the Shannon entropy of the dipole multiplicity distribution. 

As the first model for the probabilities $p_n$, one can consider a one-dimensional reduction of the DIS dipole description. For this reduced scenario, the probabilities are subject to the following evolution equation\footnote{We refer in this paper to this model as a 1+0 D case, i.e.\ 1 D in rapidity and 0 D in a transverse dimension. The general model would be the 1+2 D case and DLLA would be the 1+1 D case. This terminology differs from the one in \cite{Kharzeev:2017qzs} where they use 1+1 D and 1+3 D, respectively. In this case, they refer to underlying QFT calculations with one on-shell  condition.}: 
\begin{align} 
    \frac{dp_n(Y)}{dY} = -n\,\lambda\,p_n(Y) + (n-1) \lambda\,p_{n-1}(Y),
\label{eq:dipole0}
\end{align}
where $\lambda$ is a free parameter which can be interpreted as the BFKL intercept. 
With the initial condition $p_1(y=0) =1 $, $p_{n \geq 1}(y=0) = 0$, one has as a solution 
\begin{align}
    \label{eq:sol_1D}
    p_n(Y) & = \frac{e^{-\lambda\,Y}}{C}\left(1-\frac{e^{-\lambda Y}}{C} \right)^{n-1}, & C& =1.
\end{align}
The mean multiplicity  and entropy are  obtained as
\begin{align}
\label{eq:10reduction}
    \langle n \rangle & = \sum_n n\,p_n = C e^{\lambda Y},
    &
    S & = -\sum_n p_n \ln p_n = \left( 1-\langle n \rangle\right) \ln \left( 1- \frac{1}{\langle n \rangle}\right) + \ln \langle n \rangle.
\end{align}
Interpreting the dipole multiplicity as the total number of partons and fixing the intercept $\lambda$, and possibly also the parameter $C \neq 1$ from parton distribution functions for specific bins in the photon virtuality $Q^2$, see also \cite{Hentschinski:2022evidence} for a detailed discussion, a successful description of H1 data on hadronic final-state entropy is possible. While the so-far results confirm  the picture outlined in \cite{Kharzeev:2017qzs} and further specified above, the 1-dimensional reduced description is in general too simple. In particular, the model requires a fit of its parameter to parton distribution functions, from which the dependence on the photon virtuality is extracted, in some cases including the additional parameter $C \neq 1$. To overcome this limitation, it is therefore needed to determine probability densities $P_n$ with an explicit dependence on the size of the $n$ dipoles, but with the dependence on the momentum fraction integrated out, 
\begin{align}
  P_n(Y= \ln 1/x; \rt_1, \rt_2, \ldots, \rt_n) & =   
    \int\limits_x^1 \frac{dz_n}{z_n}  
      \ldots  \int\limits_{z_2}^1 \frac{dz_1}{z_1}
    \sum_{i_1\dots i_{n}} \sum_{j_1\dots j_{n}}  \left|a^{i_1\dots i_{n}}_{j_1\dots j_{n}}\left( \{z_l, \rt_l\}\right) \right|^2,
\end{align}
where we made the ordering of the momentum fractions explicit, $1 > z_1 > \ldots > z_n > x$, as appropriate for the leading logarithmic approximation.
 Therefore, $P_n$ denotes the probability density to encounter a system of  $n$ % {\color{red} daughter} 
dipoles at transverse separations $\rt_1$, \ldots, $\rt_n$ in the DIS reaction with $Y = \ln 1/x$. For later reference we note that the probability densities satisfy the normalization condition,
\begin{align}
    \label{eq:probs_dipole}
\sum_{n=1}^\infty  \prod_{i=1}^n \int d^2\rt_i  P_n(Y; \rt_1, \rt_2, \ldots, \rt_n) & =1,
\end{align}
 which is a direct consequence of the normalized of the density operator Eq.~\eqref{eq:trace_color}.
 
\subsection{Levin--Lublinsky equation}
\label{sec:LevinLublinsky}

Within the leading logarithmic approximation, which aims at a resummation of terms $\left(\alpha_s \ln 1/x \right)^n$ to all orders in the strong coupling constant $\alpha_s$, the probability densities $P_n$ are determined as the solution to certain evolution equations. 
The original formulation of the dipole model, presented in \cite{Mueller:1993rr}, provides a description which allows to obtain the dipole multiplicity from a certain generating functional, which itself can be obtained as the solution to a non-linear evolution equation. This framework has been explored previously in the literature, see in particular \cite{Salam:1995zd,Liou:2016mfr,Liu:2022bru}, 
using the assumption of large dipole multiplicities; see also developments addressing saturation in \cite{Salam:1995uy,Avsar:2005iz,Flensburg:2011kk}. An alternative description for obtaining dipole multiplicities was presented in  \cite{Levin:2003nc}, see also \cite{Kharzeev:2017qzs}, which provides a linear evolution equation, directly  in terms of the probability densities $P_n$. It is this framework which we will explore in the following. The evolution equation reads
\begin{align}
  \frac{dP_n(Y; \rt_1, \ldots, \rt_n)}{dY}
  = & - \sum_{i=1}^n \sigma(\rt_i)\, P_n(Y;\rt_1, \ldots, \rt_n) \nonumber\\
  & + \sum_{i=1}^{n-1} K(\rt_i, \rt_n)\, P_{n-1}(Y; \rt_1, \ldots, \rt_i + \rt_n, \ldots, \rt_{n-1}),
  \label{eq:dpn}
\end{align}
where
\begin{align}
  \label{eq:kernel}
  K(\rt_i, \rt_n) &= \frac{\asb}{2\pi} 
  \frac{(\rt_i + \rt_n)^2}{\rt_i^2 \, \rt_n^2} \,
  \theta(\rt_i^2 - \rho^2)\, \theta(\rt_n^2 - \rho^2),
  &
  \asb & = \frac{\alpha_s N_c}{\pi},
\end{align}
is the dipole splitting  kernel supplemented by an UV regulator $\rho$. It denotes the probability for a dipole of size  $(\rt_{i}+\rt_{n})$ to decay into two dipoles of sizes $\rt_{i}$ and $\rt_{n}$. The evolution equation assumes that the dipoles are homogeneously distributed and the impact-parameter dependence factorizes and can be accounted for in the initial condition \cite{Levin:2003nc}. Unlike the evolution equation in the 1+0 model,  %the color dipole amplitude, 
the above evolution equation depends on the UV regulator $\rho \to 0$.  The solution of the above system of equations requires further an infra-red (IR) cut-off $\sim 1/\Lambda_{\rm QCD}$ to prevent the formation of large dipoles.
Within a more complete description, the dipole probabilities densities still need to be convoluted with the wave function of external scattering particles. These wave functions provide then finite support only in a limited region of dipole sizes, roughly $1/\Lambda_{\rm QCD} > |\rt_i| > 1/Q$. Since our current study does not include those wave functions, we take these effects into account through setting $\rho = 1/Q$ in the following and through using a corresponding infrared cut-off, which we identify with the size of the initial dipole.

The virtual correction or the dipole survival probability can be obtained as an integral over the real emission kernel:
\begin{align}
  \label{eq:integraged_kernel}
  \sigma(\rt_i) & = \int d^2 \rt'K(\rt_i-\rt', \rt') 
  %\Theta({\rt'}^2- \rho) 
  \approx \overline{\alpha}_s\ln \frac{\rt_i^2}{\rho^2}.
\end{align}
 As the initial conditions we chose in the following:
\begin{align}
    P_n(Y=0; \rt_1, \ldots, \rt_n) & = \delta_{n1} \delta^{(2)}(\rt_1 -\rt),
\end{align}
i.e. the proton is taken to consist of a single dipole of size $\rt$ at $Y = 0$. The mean dipole multiplicity is finally defined as
\begin{align}
    \label{eq:mean_dipole}
\langle n\rangle(x,r^2/\rho) &=  \sum_{n=1}^\infty  n  \prod_{i=1}^n \int d^2\rt_i  P_n(Y; \rt_1, \rt_2, \ldots, \rt_n),
\end{align}
where the dependence of the $P_n$ on the initial conditions and cut-offs is understood.

\subsection{Dipole number distribution in DLLA} %the double logarithmic approximation}

As a special case of the above evolution equation, we will further study the dipole cascade in the double logarithmic approximation (DLLA), where we keep only track of those configurations which provide for the mean dipole number a contribution which is for each power in $\asb$ associated with a corresponding factor of $Y$ and a logarithm $\ln(\rt^2/\rho^2)$. While the dipole cascade has been already studied in the DLLA in the literature \cite{Liou:2016mfr,Liu:2022bru,Liu:2022hto},  we will revisit the problem below, starting from DLLA of the Levin--Lublinsky equation, which is on its own a new result. DLLA introduces ordering in rapidity and transverse sizes of subsequently emitted gluons.
Within the DLLA, the kernel of the evolution equation gets simplified. The probability distributions will lose information on the transverse orientation of dipoles, and their sizes will be strongly ordered $\rho \ll r_n \ll \ldots \ll r$. 
Assuming as the initial condition 
\begin{align}
    \label{eq:initialDLL}
    P^{\rm DLLA}_n(Y=0, r_1) &= \delta_{n1} \delta(r^2 - r_1^2),
\end{align}
we have
\begin{align}
    \label{eq:probs_dipoleDL}
\sum_{n=1}^\infty  \int_{\rho^2}^{r^2} dr_1^2  \int_{\rho^2}^{r_1^2} dr_2^2  \ldots  \int_{\rho^2}^{r_{n-1}^2} dr_n^2 P^{\rm DLLA}_n(Y; r_1, r_2, \ldots, r_n) & =1.
\end{align}
During the splitting process, we need now to assume that each dipole $r_i$ splits into two dipoles, where one dipole has within double-leading-logarithmic accuracy the size $r_i$ and the second dipole the size $r_{n+1}$. 
%Note that either the first or the second new dipole can be the smallest one. 
We have 
\begin{align}
  \label{eq:DLkernel}
   K^{\rm DLLA}(\rt_{n+1}, \rt) & = \frac{\overline{\alpha}_s}{\rt_{n+1}^2}  \Theta(\rt_{n+1}^2- {\rho^2})\Theta(\rt^2 - \rt_{n+1}^2).
\end{align}
Note that the above expression accounts for a possibility that any of two daughter dipoles can be the smallest one, i.e.\ it corresponds to the configuration $r_i^2 \ll r_n^2$ and $r_n^2 \ll r_i^2$ of Eq.~\eqref{eq:kernel}. The result is in accordance with both BK, BFKL, and DGLAP  evolution in the double-leading-logarithmic limit.   

To formulate the evolution equation, one starts with
\begin{align}
  \label{eq:dpn_DL1}
  \frac{dP^{\rm DLLA}_n}{dY}
 & = - \sum_{i=1}^n \sigma(\rt_n)\, P^{\rm DLLA}_n(Y; r_1, \ldots, r_n)
  + \sum_{i=1}^{n-1} K(r_{n}, r_i)\, P^{\rm DLLA}_{n-1}(Y; r_1, \ldots r_{n-1})
  \notag \\
  & 
   = - n \sigma^{\rm DLLA}(\rt_n)\, P^{\rm DLLA}_n(Y; r_1, \ldots, r_n)
  + (n-1) \frac{\overline{\alpha}_s}{r_{n}^2}  \, P^{\rm DLLA}_{n-1}(Y; r_1, \ldots r_{n-1}),
\end{align}
with a still undetermined non-emission probability $\sigma(r_n)$. Taking then the $Y$ derivative of the normalization condition of Eq.~\eqref{eq:probs_dipoleDL} fixes then
\begin{equation}
  \sigma^{\rm DLLA}(\rt_n)   =\int_{\rho^2}^{r_{n}^2} dr_{n+1}^2  \frac{\asb}{r_{n+1}^2}= \asb \ln \frac{r_n^2}{\rho^2}.
\end{equation}
Within the DLLA, it is then possible to obtain a  closed equation for the mean multiplicity of dipoles\footnote{In the context of Quantum Information, the quantity $\langle n\rangle=\sum_{n=0} p_n\,n\equiv C_K$ considered in the Krylov basis is called the Krylov complexity or the spread complexity \cite{Caputa:2024xkp,Baiguera:2025dkc}. It measures how the state spreads in a minimal basis under unitary evolution. In \cite{Caputa:2024xkp} it was demonstrated that if one evolves a state using a boost-like Hamiltonian constructed out of generators of the SL(2,R) group, the resulting probabilities obtained from squaring an absolute value of the corresponding wave function obey Eq.~(\ref{eq:dipole0}). Generalizing this, we conjecture that in the present scenario, the mean multiplicity, i.e.\ the gluon density, measures the complexity of the dipole cascade and therefore of the proton, accounting for additional degrees of freedom.}.  
With 
\begin{align}
    \label{eq:mean_dipoleDL}
\langle n\rangle\left(x, \frac{r^2}{\rho^2}\right) &= \sum_{n=1}^\infty  n  \int_{\rho^2}^{r^2} dr_1^2  \int_{\rho^2}^{r_1^2} dr_2^2  \ldots  \int_{\rho^2}^{r_{n-1}^2} dr_n^2 P_n(Y; r_1, r_2, \ldots, r_n) , 
%\notag \\
%&= 
%\sum_{n=1}^\infty  n \prod_{i=1}^n \int_{\rho^2}^{r^2} dr^2_i \theta(r_1-r_2)\theta(r_2-r_3)\ldots \theta(r_{n-1} - r_n) P_n(Y, r; r_1, r_2, \ldots, r_n)  
%\notag \\
%&= \sum_{n=1}^\infty  n  \int_{\rho^2}^{r^2} dr_n^2  \int_{r_n^2}^{r^2} dr_{n-1}^2  \ldots  \int_{r_2^2}^{r^2} dr_1^2 P_n(Y, r; r_1, r_2, \ldots, r_n)  .
\end{align}
we have 
\begin{align}
    \frac{d\langle n\rangle\left(x, \frac{r^2}{\rho^2}\right)}{dY} &= %sum_{n=1}^\infty  n  \int_{\rho^2}^{r^2} dr_1^2  \int_{\rho^2}^{r_1^2} dr_2^2  \ldots  \int_{\rho^2}^{r_{n-1}^2} dr_n^2 \frac{d P_n(Y, r; r_1, r_2, \ldots, r_n)  }{dY}
%\notag \\
%&= 
%\sum_{n=1}^\infty    \int_{\rho^2}^{r^2} dr_1^2  \int_{\rho^2}^{r_1^2} dr_2^2  \ldots  \int_{\rho^2}^{r_{n-1}^2} dr_n^2 \left[-n^2 \sigma(r_n) + \asb (n+1)n \int_{\rho^2}^{r_{n}^2} \frac{dr^2_{n+1}}{r_{n+1}^2} \right]P_n(Y, r; r_1, r_2, \ldots, r_n)
%\notag \\
%&= 
% \asb\sum_{n=1}^\infty  n  \int_{\rho^2}^{r^2} dr_1^2  \int_{\rho^2}^{r_1^2} dr_2^2  \ldots  \int_{\rho^2}^{r_{n-1}^2} dr_n^2   \int_{\rho^2}^{r_{n}^2}dr^2_{n+1} \frac{1}{r_{n+1}^2} P_n(Y, r; r_1, r_2, \ldots, r_n)
%\end{align}
%While we maintain now the ordering, we switch now the order in which the integrals are written and arrive at
%\begin{align}
%      \frac{dn(x, \rho^{-2})}{dY} &=
%      \asb\sum_{n=1}^\infty  n  
%      \int_{\rho^2}^{r^2} dr_{n+1}^2  \int_{r_{n+1}^2}^{r^2} dr_n^2  \ldots  \int_{r_2^2}^{r^2} dr_1^2    \frac{1}{r_{n+1}^2} P_n(Y, r; r_1, r_2, \ldots, r_n)
%      \notag \\
%      &= \asb\sum_{n=1}^\infty  n  
%      \int_{\rho^2}^{r^2} dr_{a}^2  \frac{1}{r_a^2} \int_{r_{a}^2}^{r^2} dr_n^2  \ldots  \int_{r_2^2}^{r^2} dr_1^2    \frac{1}{r_{n+1}^2} P_n(Y, r; r_1, r_2, \ldots, r_n)
      \asb \int_{\rho^2}^{r^2} dr_{a}^2  \frac{1}{r_a^2} \langle n\rangle\left(x, \frac{r^2}{r_a^2}\right),
\end{align}
which can be rewritten as
\begin{align}
   \frac{d^2 \langle n\rangle\left(x, \frac{r^2}{\rho^2}\right)}{dY d \ln 1/\rho^2} &=  - \rho^2 \frac{d^2\langle n\rangle\left(x, \frac{r^2}{\rho^2}\right)}{dY d \rho^2} = \asb \langle n\rangle\left(x, \frac{r^2}{\rho^2}\right).
\end{align}
This is the common double-leading-logarithmic limit of the fixed coupling DGLAP and BFKL equations which is solved by 
\begin{align}
     \langle n\rangle\left(Y, \ln \frac{r^2}{ \rho^2} \right)  & = \sum_{k=0}^\infty \frac{(\asb  \ln \frac{r^2}{ \rho^2} Y)^k}{(k!)^2}
 = I_0 \left(2\sqrt{ \asb \ln \frac{r^2}{ \rho^2} Y} \right),
 \label{eq:BesselI0}
 \end{align} 
where $I_0(x)$ is the modified Bessel function of the first kind.

\section{Multiplicity, mean multiplicity and entropy}
%This allows to determine $\Delta$ from i.e. the gluon distribution, which present a similar growth at low $x$ if $y = \ln 1/x$. 

In this section, we present numerical results for multiplicity, its mean value and entropy of dipoles generated by 
the Monte Carlo generator {\sf DIPMAR}~\cite{DIPMAR}\footnote{For other Monte Carlo solutions of the Mueller cascade equation see \cite{Salam:1996nb,Flensburg:2011kk,Domine:2018myf}.} described in detail in Appendix~B1. 
It employs the so-called Markov Chain Monte Carlo (MCMC) algorithm to solve numerically the Levin--Lublinsky equation exactly 
(up to statistical errors) as well as in the DLLA.
It was positively cross-checked with an independent Monte Carlo program {\sf LLMC} \cite{LLMC}, see Appendices B.2 and B.3.

\begin{figure}[!ht]
    \centering     
    \includegraphics[width=1.0\linewidth]{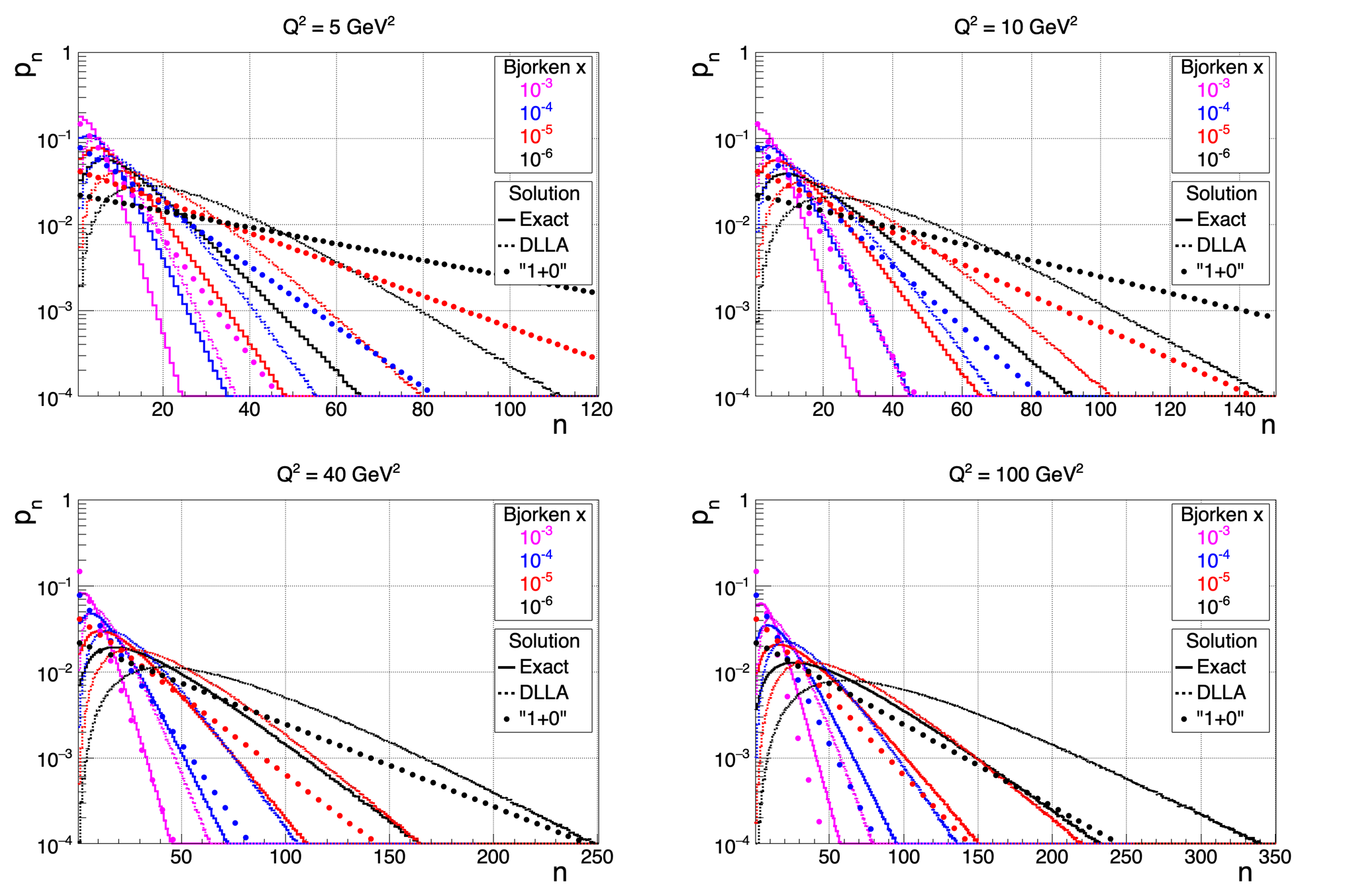}
    \caption{Monte Carlo results from {\sf DIPMAR} for probability distributions $p_n$ of a number of generated dipoles $n$
     corresponding to the fixed $\asb = 0.1$ and the initial dipole size (for `Exact' and `DLLA') 
     $r=0.85\,$fm.
     In the `Exact' case, the IR cut-off $\Delta_{\rm IR} = r = 0.85\,$fm was also applied.
     }
    \label{fig:Pn}
\end{figure}

\begin{figure}[!ht]
    \centering     
    \includegraphics[width=1.0\linewidth]{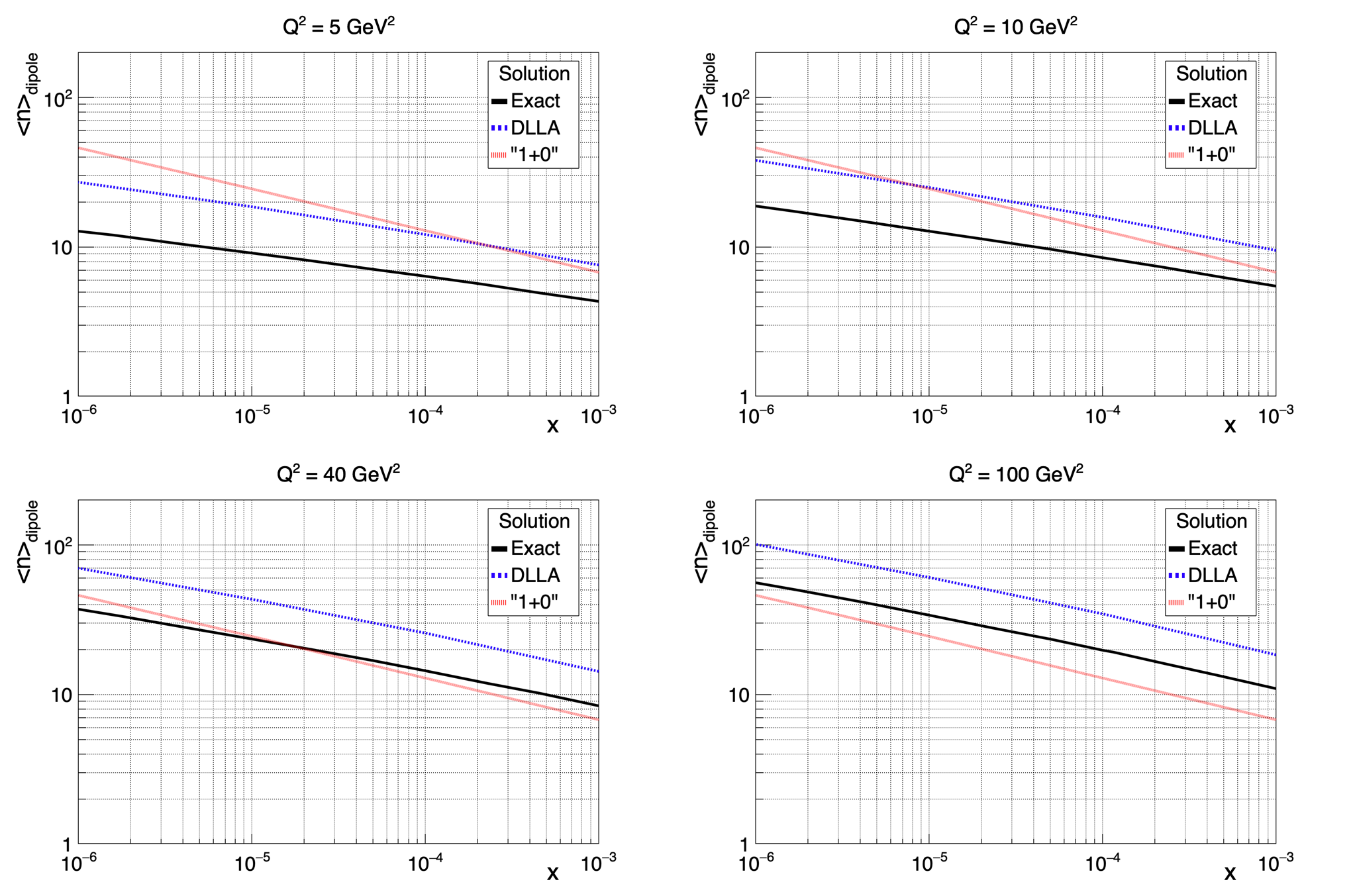}
    \caption{Monte Carlo results from {\sf DIPMAR} for the dipole mean multiplicity $\langle n \rangle_{\rm dipole}$ 
     corresponding to the fixed $\asb = 0.1$ and the initial dipole size (for `Exact' and `DLLA') $r=0.85\,$fm.
     In the `Exact' case, the IR cut-off $\Delta_{\rm IR} = r = 0.85\,$fm was also applied.
    }
    \label{fig:nmeandipole}
\end{figure}

%%%%
In Fig.~\ref{fig:Pn}, we present the results for the  multiplicity distributions integrated over all transverse dimensions 
corresponding to the exact solution (Exact) of Eq.~(\ref{eq:dpn}),
its double leading-log approximation (DLLA) and 1+0 model of Eq.~(\ref{eq:dipole0}); 
the initial dipole size $r=0.85\,$fm applies only to the exact and DLLA solutions.
In the case of the exact solution, we also applied the IR dipole cut-off $\Delta_{\rm IR} = r = 0.85\,$fm.
We see that both the exact solution and DLLA have characteristic turn over at small multiplicity values and that DLLA gives overall wider distribution than the exact solution. This is because in the case of the approximated equation, there is no requirement that the emitted dipoles should add vectorially to the parent dipole; see Appendix~B.1.3 for details. As a result, the phase space for emissions is larger and more dipoles are produced.
%In case of LL equation and the DLLA the scale dependence is determined by inverse of cutoff $\rho$. In the case of 0+1 scenario there is no scale dependence simply because there so transverse  dimension and divergence to regulate with the cutoff. 
For small values of $Q^2$, the multiplicity distribution from the 1+0 model is wider than the one obtained from the DLLA equation, and for low and moderate values of $Q^2$ even larger than the distribution obtained from the exact solution\footnote{Please note that the scales on the $x$ axis are different, the results for 1+0 scenario are the same for all the $Q^2$ values.}.  
Some features of these differences can be better analyzed for mean multiplicity $\langle n \rangle (x,Q^2)$ shown in Fig.~\ref{fig:nmeandipole}.  
We find that the exact result for the mean multiplicity lies below the  DLLA  result. %; and 1+0 is the smallest in the whole $x$ range only for $Q^2=100\,$GeV$^2$.

%Another quantity that can be calculated from the obtained solutions is the dipole entropy. 
%In \cite{Kharzeev:2017qzs}, this entropy has been conjectured to be an entanglement entropy that follows from incomplete measurement of the proton's wave function \cite{Kharzeev:2017qzs,Kharzeev:2021nzh,Hentschinski:2024qcd_evo,Kharzeev:2026inq} which is understood as a measurement with some resolution scales both in the longitudinal and transverse directions, averaged over color.
We finally study the  Shannon entropy of the multiplicity of dipoles
\begin{equation}
    S = - \mathrm{Tr}\,\big(\rho \ln \rho \big)=-\sum_{n=1}^{\infty} p_n\ln p_n\,.
    \label{eq:Sent}
\end{equation}
%where $p_n$ is given in Eq.~(\ref{eq:pndef}).
\begin{figure}[!ht]
    \centering     
    \includegraphics[width=1.0\linewidth]{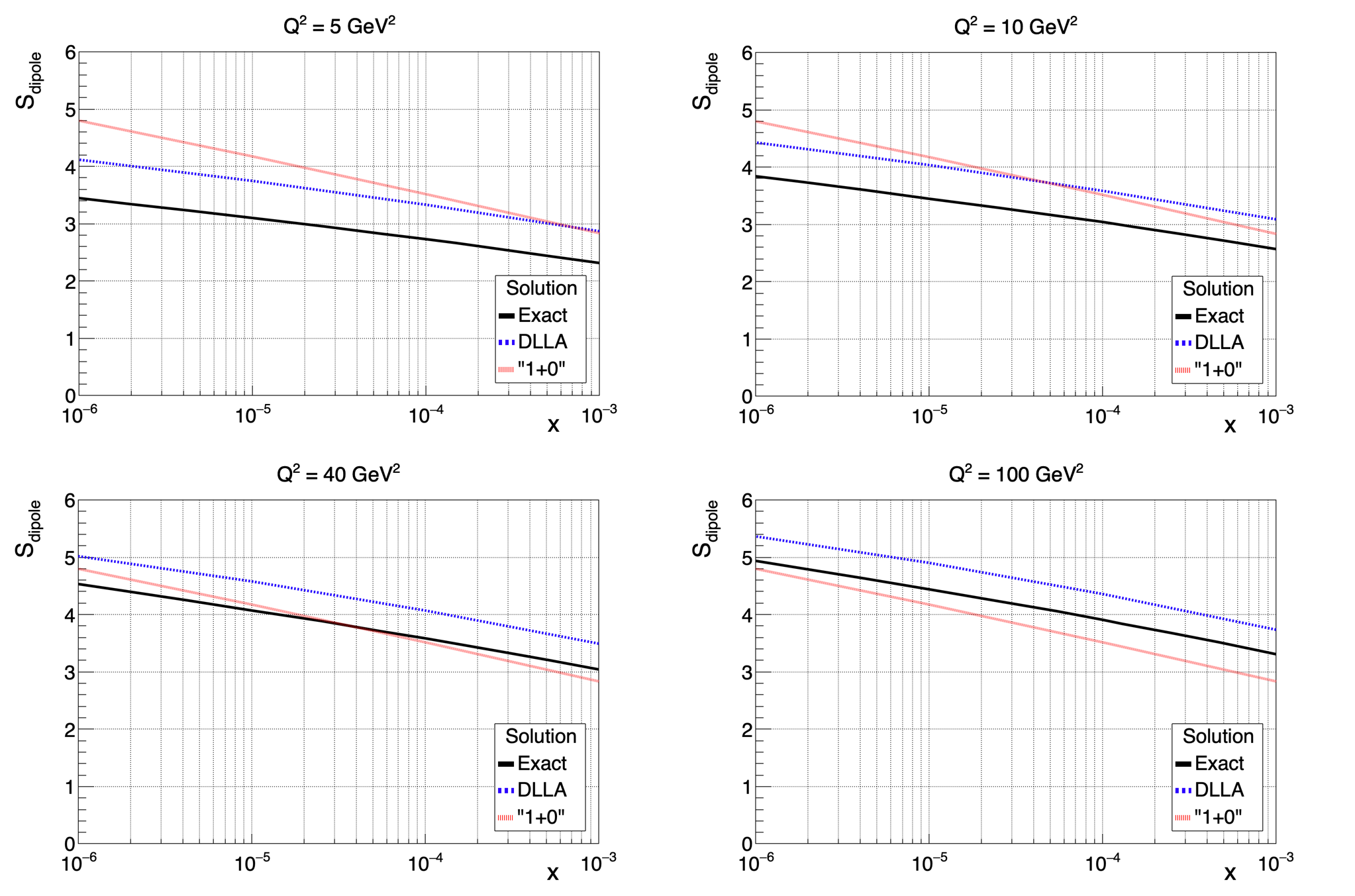}
    \caption{Monte Carlo results from {\sf DIPMAR} for the dipole entropy $S_{\rm dipole}$ 
     corresponding to the fixed $\asb = 0.1$ and the initial dipole size (for `Exact' and `DLLA') $r=0.85\,$fm.
     In the `Exact' case, the IR cut-off $\Delta_{\rm IR} = r = 0.85\,$fm was also applied.
}
    \label{fig:Sdipole}
\end{figure}
%It measures how much information has been lost due to integrating out unobservable degrees of freedom, which are phases, colours, and transverse coordinates, except for the initial dipole size.
The results for the considered cascades are presented in Fig.~\ref{fig:Sdipole}. We see that the curves follow similar patterns as the mean multiplicity distributions. However, now we can give the result additional meaning coming from information theory point of view. In the case of DLLA, the entropy is larger because, as discussed above, the information about sizes of daughter dipoles is lost because the vectorial sum of daughter dipoles is not constrained by the parent dipole as it is in the case of the exact solution. In the 1+0 model, which is independent of $Q^2$ if the BFKL intercept $\lambda$ is not adjusted by hand, the information content is the same for each bin in  $Q^2$ and in some cases it is larger and in other cases smaller than that obtained from the other two equations.

%In addition to entropy, solutions to the cascade equation can be used to calculate
%various quantum information (QI) measures. In the paper \cite{} relation has been established between QI measures and the dipole model. In particular the
%mean number of dipoles n, called in the QI context a complexity, measures how the underlying quantum state spreads in the Hilbert space

\section{Entanglement entropy and DIS data}

In this section, we present the results corresponding to comparisons of the entropy calculated using the dipole cascade of 
Eq.~(\ref{eq:dpn}) to the hadron entropy as obtained from hadron multiplicities by the H1 Collaboration \cite{H1:2020zpd}. 
%While the complete determination of the resulting entanglement entropy is experimentally  challenging, it is possible to access the entropy of the hadron multiplicity distribution.  It was argued in \cite{Kharzeev:2017qzs} that the entropy of the hadronic multiplicity distribution can be compared to the entropy related to the 
i.e.\ we provide a comparison of the Shannon entropy of the 
dipole multiplicity distribution, $S = -\sum_n p_n \ln p_n$, which is the entanglement entropy in low $x$ DIS, to the measured hadronic entropy. While previous phenomenological studies \cite{Hentschinski:2022evidence,Hentschinski:2023maxent,Hentschinski:2024qcd_evo,Hentschinski:2025pyq} found good agreement between dipole entanglement entropy and hadronic entropy, the result required fitting of free parameters of the model to parton distribution functions for each bin in $Q^2$.
%where
%\begin{align}
%    p_n(Y) &= \prod_{i=1}^n \int d^2\rt_i  P_n(Y; \rt_1, \rt_2, \ldots, \rt_n).
%\end{align}

The multiplicities as obtained from Eq.~(\ref{eq:dpn}) are dipole multiplicities, while the detectors measure only charged hadrons. In \cite{Hentschinski:2022evidence} this was taken into account by rescaling the mean multiplicity,
\begin{equation}
    \langle n \rangle \rightarrow \frac{2}{3}\langle n\rangle.
\end{equation}
This, however, cannot be done directly if one calculates probabilities from evolution equations as done here. If $P_n$ is the probability to encounter $n$ dipoles and if we assume that each dipole can associated with the production of a hadaron, then $P_N$ -- the probability to have $N$ charged pions -- can be obtained through
\begin{align}
    P_N & = \sum_{n=N}^\infty\binom{n}{N} q^N (1-q)^{n-N} P_n, \qquad q = \frac{2}{3}.
    \label{eq:PN}
\end{align}

\begin{figure}[!ht]
    \centering     
    \includegraphics[width=1.0\linewidth]{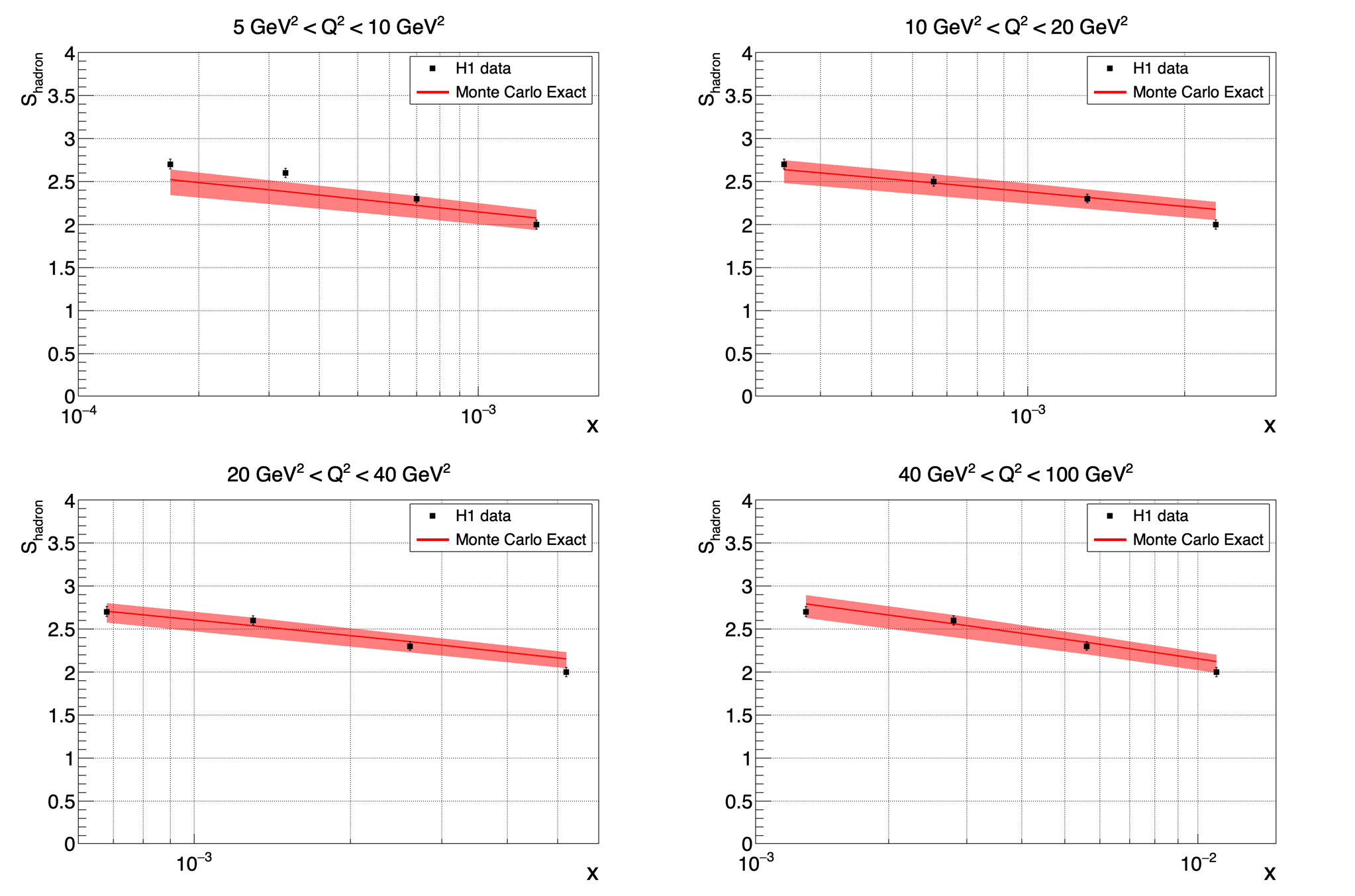}
    \caption{Comparison of Monte Carlo results from {\sf DIPMAR} with the H1 data for the hadron entropy $S_{\rm hadron}$. 
     The numerical results have been obtained through the MCMC solution of the exact dipole evolution equation with 
     the fixed $\asb = 0.1$ and the initial dipole size $r=1.0\,$fm (with $\Delta_{\rm IR} = r$). 
     The uncertainty bands correspond to the $Q^2$ ranges of the H1 measurements.}
    \label{fig:entrfull}
\end{figure}

Using the above expression  and the solution of Eq.~(\ref{eq:dpn}),
we can calculate the hadronic entropy that is directly related to the entropy of dipoles. In the calculations we used for the initial dipole size $r=1 fm$, which is quite close to the physical size of proton, while we employ for the strong coupling constant the value $\bar\alpha_s=0.1$.

Even though this value is significantly smaller than what one would expect from the given photon virtualities, a small running coupling constant is needed to reproduce with the fixed coupling LO BFKL evolution an effective Pomeron intercept of the order of $0.2\,$--$\,0.35$ as observed by the HERA experiments; the articifially low value for $\bar\alpha_s$ originates therefore in the properties of the BFKL evolution equation and not in the particular implementation of the dipole cascade in the present work. Indeed, resummation of  NLO corrections to the BFKL kernel, which are enhanced by the first coefficient of the QCD $\beta$ function, into the running coupling is known to give rise to a significantly smaller running coupling constant for the BFKL kernel than one would expect from taking into account the virtualities of external particles only \cite{Brodsky:1998kn}. In \cite{Hentschinski:2012kr,Hentschinski:2013id}, such a prescription was used for a successful NLO BKFL fit to combined HERA data, which then has been applied to a description of the LHC data in \cite{Bautista:2016xnp,ArroyoGarcia:2019cfl,Hentschinski:2025ovo}. The choice of $\bar\alpha_s=0.1$ is therefore to be understood in light of those considerations and  provides a simple but effective way to take into account important NLO corrections within a fixed-coupling LO setup. 

Our results for the hadron entropy are presented in Fig.~\ref{fig:entrfull}. Within the 1+0 dimensional reduction of the dipole cascade, see Eq.~\eqref{eq:10reduction}, entanglement entropy can be expressed as 
\begin{equation}
% WP:   S=\ln\left(\frac{2}{3} \langle n\rangle\right)+C.
   S=\ln\langle n\rangle+C,
   \label{eq:SmmC}
\end{equation}
in the limit of large dipole multiplicities. While the 1+0 dimensional reduction yields $C =1$, one finds within the double logarithmic approximation of \cite{Liu:2022bru} a value of $C \simeq 0.724$. It is then this limit of entanglement entropy which is associated with maximal entanglement.

\begin{figure}[!ht]
    %\centering     
    \includegraphics[width=1.0\linewidth]{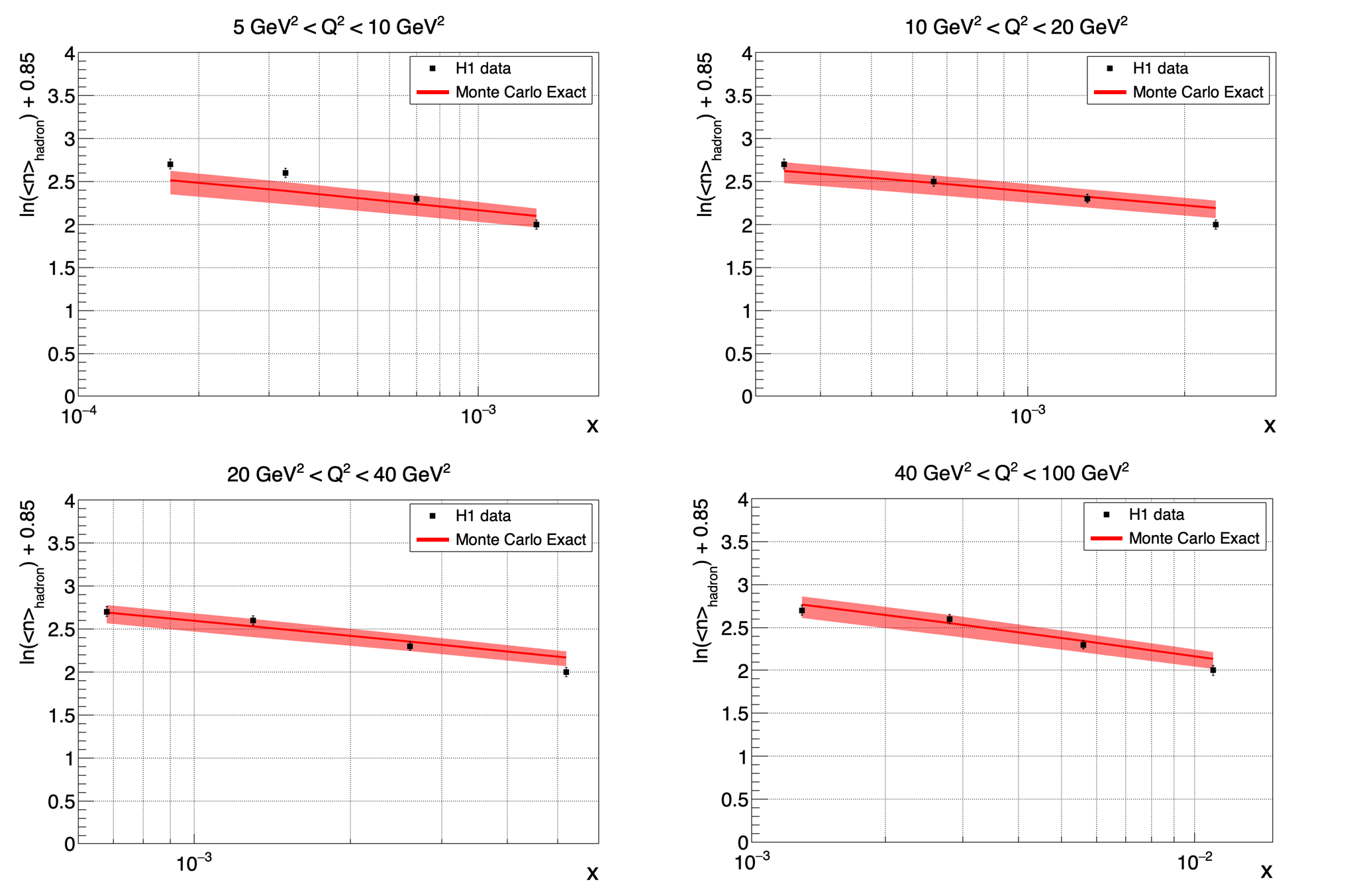}
    \caption{Comparison of Monte Carlo results from {\sf DIPMAR} with the H1 data for $\ln(\langle n \rangle_{\rm hadron}) + C$,
    where $C = 0.85$. 
     The numerical results have been obtained through the MCMC solution of the exact dipole evolution equation with 
     the fixed $\asb = 0.1$ and the initial dipole size $r=1.0\,$fm (with $\Delta_{\rm IR} = r$). 
     The uncertainty bands correspond to the $Q^2$ ranges of the H1 measurements.}
    \label{fig:entropymean}
\end{figure}

The corresponding results are presented in Fig.~\ref{fig:entropymean}. The asymptotic expression with $C=0$  undershoots the data. By the requirement to describe the data, we extract the value of $C$ which reads $C=0.85$. This value is not far from the 1+0 model where $C=1$. With this calculation, we 
also confirm that Eq.~(\ref{eq:dpn}) predicts a linear dependence of entropy on rapidity $Y=\ln(1/x)$. 

\begin{figure}[!ht]
    \centering     
    \includegraphics[width=1.0\linewidth]{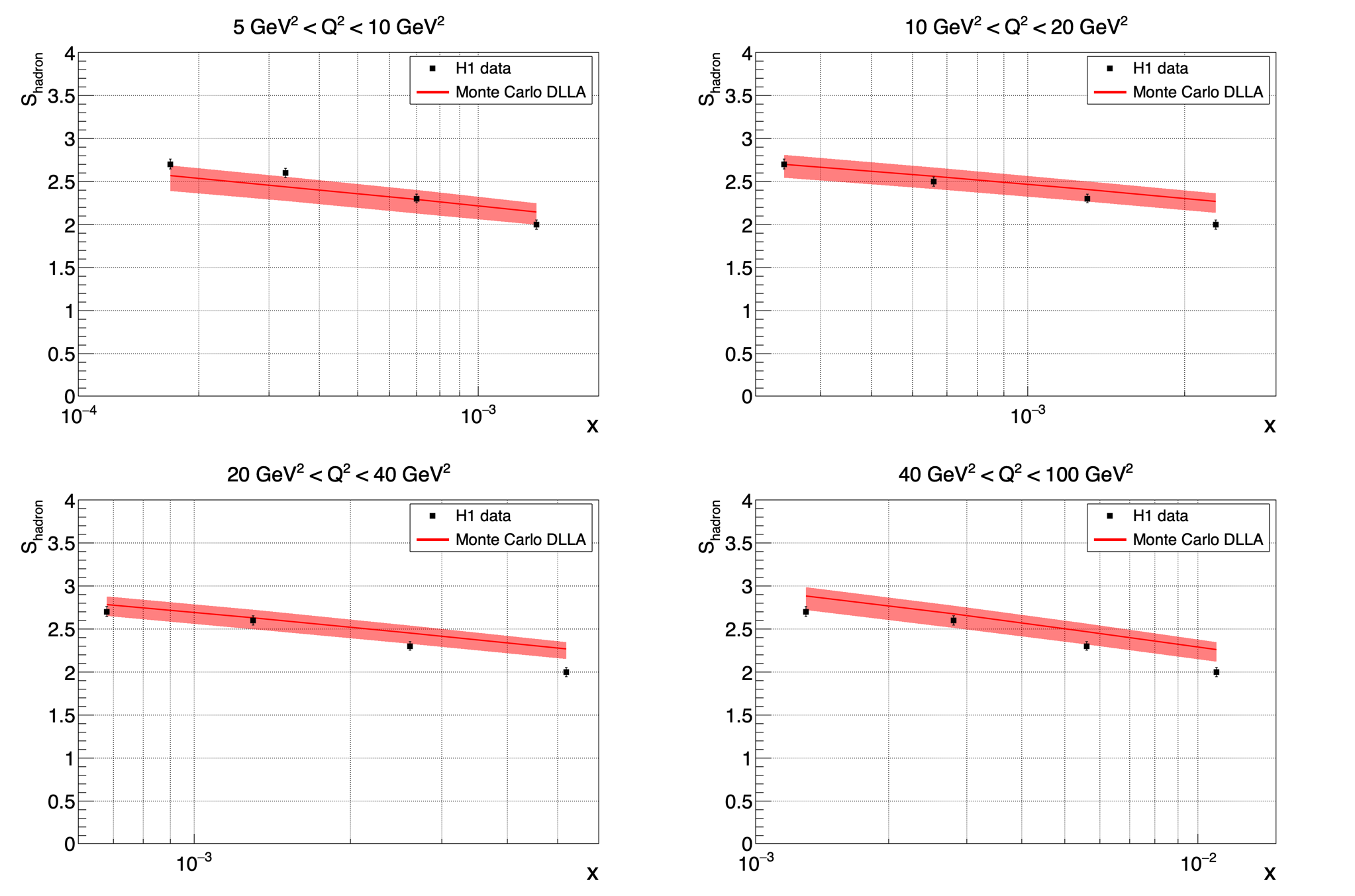}
    \caption{Comparison of Monte Carlo results from {\sf DIPMAR} with the H1 data for the hadron entropy $S_{\rm hadron}$. 
     The numerical results have been obtained through the MCMC solution of the dipole evolution equation 
     in the double leading-log approximation (DLLA) with the fixed $\asb = 0.1$ and the initial dipole size $r=0.5\,$fm. 
     The uncertainty bands correspond to the $Q^2$ ranges of the H1 measurements.}
    \label{fig:entDLLDIS}
\end{figure}

%Now we proceed to the description of data using the DLL cascade. 
In Fig.~\ref{fig:entDLLDIS}, we present the results for the hadron entropy corresponding to the DLLA cascade. While in this case one can also describe the hadron entropy data, the value used for the initial size of the dipole $r=0.5\,$fm is by a factor of $2$ smaller than in the exact solution. Thus, from this we conclude that the full cascade where the dipoles transverse sizes are not strictly ordered is the preferred physical scenario.

\begin{figure}[!ht]
    %\centering     
    \includegraphics[width=1.0\linewidth]{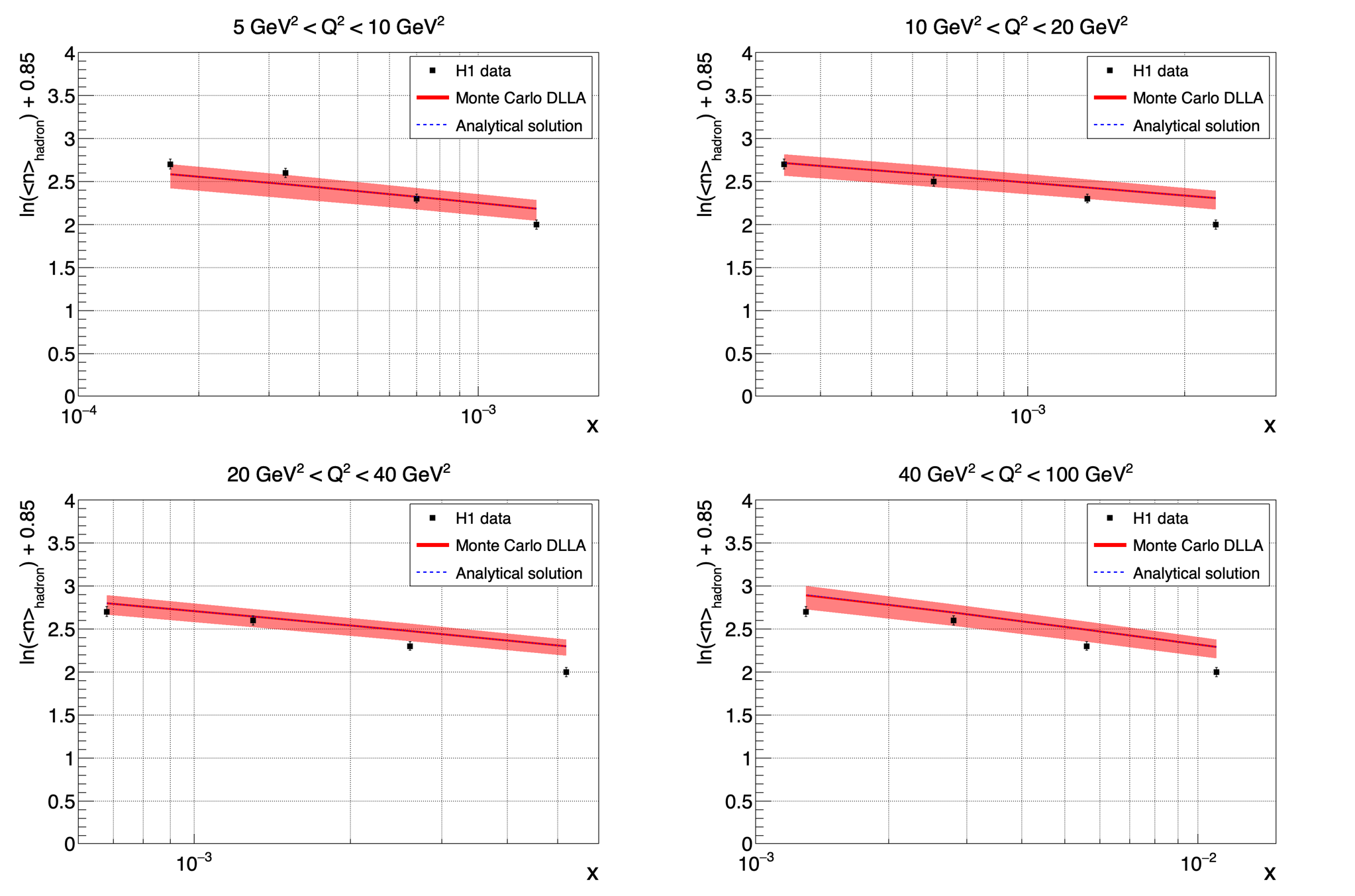}
    \caption{Comparison of Monte Carlo results from {\sf DIPMAR} with the H1 data for $\ln(\langle n \rangle_{\rm hadron}) + C$,
     where $C = 0.85$. 
     The numerical results have been obtained through the MCMC solution of the dipole evolution equation 
     in the double leading-log approximation (DLLA) with the fixed $\asb = 0.1$ and the initial dipole size $r=0.5\,$fm. 
     The uncertainty bands correspond to the $Q^2$ ranges of the H1 measurements.
     With the blue dashed lines we show the results of analytical solutions in DLLA for the central values of each $Q^2$ bin
     according to Eq.~(\ref{eq:SmmCDLLA}) -- they overlap with the corresponding Monte Carlo results (red solid lines).}
    \label{fig:entmeanDLLDIS}
\end{figure}

Similarly as for the exact cascade, we calculate the hadron entropy from the mean multiplicity. The corresponding results are presented in Fig.~\ref{fig:entmeanDLLDIS}. 
Here, we also show the results of the analytical solution (dashed line) according to the formula
\begin{equation}
 S=\ln\left(\frac{2}{3} \langle n\rangle\right)+C,
   \label{eq:SmmCDLLA}
\end{equation}
where the mean multiplicity $\langle n\rangle$ is calculated using Eq.~(\ref{eq:BesselI0}) for the central values of the $Q^2$ bins. We see a perfect agreement with the corresponding Monte Carlo results.
The extracted value of the parameter $C$ is the same as in the previous case, i.e.\ $C=0.85$, which suggests certain universality.

Let us add a comment here that we have attempted to describe the H1 entropy results accounting for the running QCD coupling constant, using various prescriptions for the dependence of $\bar\alpha_s$ on the scale. However, we could not describe the data with the universal size of the initial dipole, i.e.\ for each $Q^2$ value a different initial-dipole size was needed.

%It has been recognized that at the high rapidity and high hard scale the multiplicities obey certain type of scaling called KNO scaling. In particular the multiplicity can be expressed as 
%\begin{equation}
%    P_n=\frac{1}{\langle n\rangle} f\left(\frac{n}{\langle n \rangle}\right)
%\end{equation}
%where $f$ is universal function called KNO function. 
%Recently in \cite{} the KNO function has been explicitly constructed in the DLL approximation using analytical solution of the Mueller model. In our approach we go beyond analytical solution as it involves approximations that we can wave. 
%The results for the KNO scaling are presented on Figs.~\ref{fig:KNOr} and~\ref{fig:KNOalpha}  where our result is compared to the one obtained in \cite{}. 

%\begin{figure}
%    \centering
%    \includegraphics[width=0.5\linewidth]{dla/kno_y_dist_alpha0p2_rvar_dla.pdf}
%    \caption{KNO-scaling function for sets of dipoles with  $\bar{\alpha}_s=0.2$ for different length of the initial dipole $r$ as indicated.}
%    \label{fig:KNOr}
%\end{figure}

%\begin{figure}
%    \centering
   % \includegraphics[width=0.5\linewidth]{dla/kno_y_dist_alphavar_r3_dla.pdf}
%    \caption{KNO-scaling function for sets of dipoles with  $r=3$ for different values of $\bar{\alpha}_s$ as indicated.}
%    \label{fig:KNOalpha}
%\end{figure}

% \begin{figure}
%     \centering
%     \includegraphics[width=0.45\linewidth]{dla/kno_y_dist_alpha0p1_r2_dla.pdf} 
%      \includegraphics[width=0.45\linewidth]{dla/kno_y_dist_alpha0p2_rvar_dla.pdf} 
%     \caption{}
%     \label{fig:pn1D}
% \end{figure}

\newpage
\section{Conclusions}

In this paper, we have explored further the proposal of \cite{Kharzeev:2017qzs} that entropy generated in the DIS reaction is entanglement entropy, which arises due to the interaction of the virtual photon with the observed proton. In support of this proposal we have presented in this paper an argument which relates unobserved color degrees of freedom of the color dipole system to the emergence of a mixed state, diagonal in dipole number and with finite entanglement entropy. Even though the assumed average over color dipole degrees of freedom might be too simple in general, it nevertheless provides an intuitive explanation for the emergence of a system which can be characterized by the number of observed hadrons, which we assume to be closely related to the number of dipoles. 

Within a leading logarithmic approximation, the elements of this density matrix can then be obtained as a solution of certain evolution equations, which we study both for the complete BFKL cascade and its double logarithmic approximation. 
%In this paper, we have studied entropy production in DIS following the proposal of \cite{Kharzeev:2017qzs}. 
To this end, we have developed new Monte Carlo algorithms and implemented the corresponding computer codes, which enable the relevant calculations. 
%%%% I think it would be good to give here another statmeent about the codes, since this is central to our work
From these calculations, we obtained multiplicity distributions and used them to evaluate the entropy. By choosing phenomenologically motivated values for the initial dipole size, which models the initial size of the proton, and for the QCD coupling constant, motivated by BFKL fits, we were able to describe the entropy data extracted from the H1 analysis of hadron multiplicities.

On the theoretical side, we have demonstrated that the full cascade predicts entropy growth consistent with earlier results and exhibits, to a good approximation, a linear dependence on rapidity \cite{Kutak:2011rb}\footnote{See https://arxiv.org/abs/1103.3654v1.}, \cite{Kharzeev:2017qzs,Gursoy:2023hge}. This growth is consistent with the maximal-entanglement hypothesis \cite{Kharzeev:2026jkq}, up to an additive constant of order unity. The exact value of this constant, extracted by comparing the entropy obtained under the maximal-entanglement assumption with the direct evaluation of the Shannon entropy, is $C=0.81$.

In addition, we have obtained results in the double-leading-logarithmic approximation (DLLA). The comparison between the DLLA and the complete calculation shows that the approximate cascade contains less information, as reflected in its larger entropy. Moreover, its growth with rapidity deviates from the linear behavior. Overall, we conclude that
\begin{itemize}

\item the dipole cascade equation, Eq.(\ref{eq:dpn}), indeed describes correctly entropy that is generated  from partonic color degrees of freedom to colorless hadrons as measured by the H1 collaboration;

\item the relation between entropy and the gluon density, which in the present formulation is expressed either through the dipole multiplicity or, in the language of quantum information, through complexity,
$
S=\ln\left(\frac{2}{3} xg(x,Q^2)\right)+C ,
$
provides a viable proposal for connecting the partonic momentum density with the entropy of produced hadrons. However, its phenomenological application must take into account that experimentally only charged hadrons are measured, as reflected by the factor of $2/3$.

\end{itemize}

The obtained results support and confirm earlier findings \cite{Hentschinski:2022evidence,Hentschinski:2024qcd_evo}, based on phenomenological models and Monte Carlo simulations \cite{Hentschinski:2025pyq}, indicating that a large amount of entropy is generated already in the initial stages of the collision. This raises important questions about the possible implications for Monte Carlo simulations of proton--proton collisions, in particular those employing fragmentation functions \cite{Datta:2024hpn}, as well as for onset of hydrodynamical behavior and the thermalization of the quark--gluon plasma \cite{Janik:2025bbz,Florio:2025hoc}.

\section*{Acknowledgments}
We would like to thank Dmitri~Kharzeev, Zhoudunming Kong Tu and Stéphane Munier for informative discussions. 
The research of WP has been supported in part by a grant from the Priority Research Area (DigiWorld) under the Strategic Programme Excellence Initiative at the Jagiellonian University in Krakow, Poland. 

\appendix

\section{DLLA}

In this Appendix we demonstrate how to determine the non-splitting term and how to get the DLLA and an equation for the mean multiplicity in DLLA.

\subsection{Non-splitting term in Eq.~(\ref{eq:dpn})}

In order to obtain the expression for the non-splitting term, we start with the unitarity-condition equation for the probabilities (\ref{eq:probs_dipole}) and differentiate it on both sides as follows:
\begin{align}
    \label{eq:fix_sigma}
0&=\frac{d}{dY}\sum_{n=1}^\infty  \prod_{i=1}^n \int d^2\rt_i  P_n(Y, \rt; \rt_1, \rt_2, \ldots, \rt_n) 
\notag \\
&
= -\sum_{n=1}^\infty  \prod_{i=1}^n \int d^2\rt_i \sum_{i=1}^n \sigma(\rt_i)P_n(Y, \rt; \rt_1, \rt_2, \ldots, \rt_n) 
\notag \\
& \hspace{4cm}
+
\sum_{n=1}^\infty  \prod_{j=1}^{n+1} \int d^2\rt_j \sum_{i=1}^n K(\rt_i, \rt_{n+1})P_n(Y, \rt; \rt_1, \ldots, \rt_i + \rt_{n+1} ,\ldots\rt_n)
\notag \\
&
= -\sum_{n=1}^\infty  \prod_{i=1}^n \int d^2\rt_i \sum_{i=1}^n \sigma(\rt_i)P_n(Y, \rt; \rt_1, \rt_2, \ldots, \rt_n) 
\notag \\
& \hspace{4cm}
+
\sum_{n=1}^\infty  \prod_{j=1}^{n+1} \int d^2\rt_j \sum_{i=1}^n K(\rt_i-\rt_{n+1}, \rt_{n+1})P_n(Y, \rt; \rt_1, \ldots, \rt_i  ,\ldots\rt_n),
\end{align}
where in the last line the integral over $d^2\rt_{n+1}$ can now be carried out. We have 
\begin{align}
    \label{eq:fix_sigma2}
    0 & =  \sum_{n=1}^\infty  \prod_{i=1}^n \int d^2\rt_i \sum_{i=1}^n \left[- \sigma(\rt_i) + \int d^2 \rt_{n+1} K(\rt_i-\rt_{n+1}, \rt_{n+1})\right]P_n(Y, \rt; \rt_1, \rt_2, \ldots, \rt_n), 
\end{align}
which then fixes $\sigma(\rt_i)$. 

\subsection{Details of derivation of non-splitting term in DLLA equation for $P_n$}

Similarly as in the case of the complete cascade, we start with unitarity condition and differentiate it over rapidity:
% We have 
\begin{align}
    \label{eq:fix_sigma}
0 & = \frac{d}{dY}\sum_{n=1}^\infty  \int_{\rho^2}^{r^2} dr_1^2  \int_{\rho^2}^{r_1^2} dr_2^2  \ldots  \int_{\rho^2}^{r_{n-1}^2} dr_n^2 P_n(Y, r; r_1, r_2, \ldots, r_n)  = 
\notag \\
& =
\sum_{n=1}^\infty  (-n) \int_{\rho^2}^{r^2} dr_1^2  \int_{\rho^2}^{r_1^2} dr_2^2  \ldots  \int_{\rho^2}^{r_{n-1}^2} dr_n^2  \sigma(r_n) P_n(Y, r; r_1, r_2, \ldots, r_n)
\notag \\
& \hspace{4cm} +
\sum_{n=1}^\infty  n \int_{\rho^2}^{r^2} dr_1^2  \int_{\rho^2}^{r_1^2} dr_2^2  \ldots  \int_{\rho^2}^{r_{n}^2} dr_{n+1}^2  \frac{\asb}{r_{n+1}^2}  P_n(Y, r; r_1, r_2, \ldots, r_n)
 \notag \\
 &=
 \sum_{n=1}^\infty  n \int_{\rho^2}^{r^2} dr_1^2  \int_{\rho^2}^{r_1^2} dr_2^2  \ldots  \int_{\rho^2}^{r_{n-1}^2} dr_n^2  \left[\int_{\rho^2}^{r_{n}^2} dr_{n+1}^2  \frac{\asb}{r_{n+1}^2} - \sigma(r_n) \right] P_n(Y, r; r_1, r_2, \ldots, r_n),
\end{align}
which then fixes $\sigma(r_n) = \asb \ln (r_n^2/\rho^2)$.

%%%%%%%%%%%%%%%%%%%%%%%%%%%%%%%%%%%%%%%%%%%%%%%%%%%%%%%%%%%%%%%%%%%%%%%%%%%%%%%%%%%%%%
\section{Monte Carlo solutions of Levin--Lublinsky equation}

Here, we describe two Monte Carlo algorithms developed independently by two of us to solve the Levin--Lublinsky equation 
for dipole evolution given in Subsection~\ref{sec:LevinLublinsky}. They are implemented in two different Monte Carlo programs: 
{\sf DIPMAR}~\cite{DIPMAR} and {\sf LLMC}~\cite{LLMC}.

\subsection{Monte Carlo generator {\sf DIPMAR}}
\label{App:B1}

\subsubsection{Dipole distribution function}
\label{sssec:ddf}

Let us define a dipole distribution function $D(\xt,y;\rho)$ as the probability that the initial dipole $\rt_1$ at some initial rapidity $y_0$, 
after evolution, becomes equal to $\xt$ at rapidity $y$:
\begin{equation}
  \label{eq:Dxy}
 \begin{aligned}
  D(\xt,y;\rho) & \equiv  \sum_{n=1}^\infty \prod_{i=1}^n\int d^2 \rt_i   \delta(\xt-\rt_1) P_{n}(y, \rt_1, \ldots, \rt_n;\rho)  
\\ & =  \sum_{n=1}^\infty \prod_{i=2}^n\int d^2 \rt_i  P_{n}(y, \xt,\rt_2, \ldots, \rt_n;\rho),
\end{aligned}
 \end{equation}
where $P_{n}(y, \rt_1, \ldots, \rt_n;\rho)$ is the probability of having $n$ dipoles $\rt_1, \ldots, \rt_n$ at rapidity $y$, 
assuming the UV cut-off $\rho$, given by Eq.~(\ref{eq:dpn}).
$D(\xt,y;\rho)$ is normalized to $1$ because
\begin{equation}
 \label{eq:DdyNorm}
\int d^2\xt \, D(\xt,y;\rho)  = \sum_{n=1}^\infty \prod_{i=1}^n\int d^2 \rt_i \,  P_{n}(y, \rt_1, \ldots, \rt_n;\rho)   = \sum_{n=1}^\infty p_{n}(y;\rho) = 1.  
\end{equation}

Then, one can obtain the following differential equation:
\begin{align}
%\label{eq:dDdy1}
\nonumber
\frac{\partial D(\xt,y;\rho)}{\partial y}  =  & \sum_{n=1}^\infty \prod_{i=2}^n\int d^2 \rt_i \, \frac{\partial P_{n}(y, \xt,\rt_2, \ldots, \rt_n;\rho)}{\partial y}
\\
%\label{eq:dDdy2}
\nonumber
 = & - \sum_{n=1}^\infty \prod_{i=2}^n\int d^2 \rt_i  \left[\sigma(x;\rho) + \sum_{j=2}^n \sigma(r_j;\rho)\right] P_n(y,\xt,\rt_2, \ldots, \rt_n;\rho)
 \\  
\nonumber
%\label{eq:dDdy3}
& + \sum_{n=1}^\infty \prod_{i=2}^n\int d^2 \rt_i  \Bigg[ K(\xt, \rt_n;\rho)\, P_{n-1}(y, \xt + \rt_n, \rt_2, \ldots, \rt_{n-1};\rho) 
\\ & \hspace{25mm} + \sum_{j=2}^{n-1}
  K(\rt_j ,\rt_n;\rho)\, P_{n-1}(y, \xt, \rt_2, \ldots, \rt_j + \rt_n, \ldots, \rt_{n-1};\rho)\Bigg]
\nonumber
\\
\label{eq:dDdy}
= & -\sigma(x;\rho)D(\xt,y;\rho) + \int d^2\rt \,K(\xt,\rt;\rho)D(\xt+\rt,y;\rho), 
\end{align}
where we have used Eqs.~(\ref{eq:dpn}) and (\ref{eq:Dxy}).

Finally, we can write the differential equation for $D(\xt,y;\rho)$ in the form
\begin{equation}
  \label{eq:dDxdy}
\frac{\partial D(\xt,y;\rho)}{\partial y}  = -\sigma(x;\rho)D(\xt,y;\rho) + \int d^2\xt'  d^2\rt \,\delta(\xt+\rt-\xt') 
K(\xt'-\rt,\rt;\rho)D(\xt',y;\rho).
\end{equation}
Its formal solution is given in the form of an integral equation
\begin{equation}
  \label{eq:DxyIntEq}
  \begin{aligned}
D(\xt,y;\rho)  =\:\; & e^{-\sigma(x;\rho)(y-y_0)}D(\xt,y_0;\rho) \\
&+ \int_{y_0}^ydy' \,e^{-\sigma(x;\rho)(y-y')} \int d^2\xt'  d^2\rt \,\delta(\xt+\rt-\xt') K(\xt'-\rt,\rt;\rho)D(\xt',y';\rho),
\end{aligned}
\end{equation}
which can be checked by differentiation over $\partial y$.

The above equation can be solved by iteration. The $n$-step iterative solution reads
\begin{align}
 D_n(\xt,y;\rho)  =\: & \int d^2\xt_0\, D(\xt_0,y_0;\rho) \Bigg\{ e^{-\sigma(x;\rho)(y-y_0)}\,\delta(\xt-\xt_0) 
\nonumber\\
  &\hspace{28mm}  
  + \sum_{j=1}^n \prod_{i=1}^j \left[ \int_{y_{i-1}}^y dy_i \int d^2\rt_i \,K(\xt_i,\rt_i;\rho) \,e^{-\sigma(\xt_{i-1};\rho)(y_i-y_{i-1})}\right]
 \nonumber\\
\label{eq:DnIter}
 & \hspace{38mm} \times e^{-\sigma(\xt_j;\rho)(y-y_j)}\,\delta(\xt-\xt_j)\Bigg\},
\end{align}  
where $\xt_i = \xt_{i-1} - \rt_i$.

The final iterative solution is obtained by taking in the above the limit $n\rightarrow \infty$, i.e.
\begin{equation}
  \label{eq:DnLimit}
D(\xt,y;\rho)  = \lim_{n\rightarrow\infty} D_n(\xt,y;\rho).
\end{equation}
Thus, we get
\begin{align}
 D(\xt,y;\rho)  =\: & \int d^2\xt_0\, D(\xt_0,y_0;\rho) \Bigg\{ e^{-\sigma(x;\rho)(y-y_0)}\,\delta(\xt-\xt_0) 
\nonumber\\
  &\hspace{28mm}  
  + \sum_{n=1}^{\infty} \prod_{i=1}^n \left[ \int_{y_{i-1}}^y dy_i \int d^2\rt_i \,K(\xt_i,\rt_i;\rho) \,e^{-\sigma(\xt_{i-1};\rho)(y_i-y_{i-1})}\right]
 \nonumber\\
\label{eq:DIterSol1}
 & \hspace{38mm} \times e^{-\sigma(\xt_n;\rho)(y-y_n)}\,\delta(\xt-\xt_n)\Bigg\}.
\end{align}  

Since 
\begin{equation}
  \label{eq:ExpInt}
e^{-\sigma(x_n;\rho)(y-y_n)}  = \int_y^{\infty} dy_{n+1}\, \sigma(x_n;\rho)\,e^{-\sigma(x_n;\rho)(y_{n+1}-y_n)},
\end{equation}
the above solution can also be expressed in a form more suited for evaluation with the Markov Chain Monte Carlo (MCMC)
method:
\begin{equation}
\begin{aligned}
 D(\xt,y;\rho)  =\: & \int d^2\xt_0\, D(\xt_0,y_0;\rho) \Bigg\{ \int_y^{\infty} dy_1 \,\sigma(x_0;\rho)\,e^{-\sigma(x_0;\rho)(y_1-y_0)}\,\delta(\xt-\xt_0) 
\\
  &\hspace{28mm}  
  + \sum_{n=1}^{\infty} \prod_{i=1}^n \left[ \int_{y_{i-1}}^y dy_i \int d^2\rt_i \,K(\xt_i,\rt_i;\rho) \,e^{-\sigma(\xt_{i-1};\rho)(y_i-y_{i-1})}\right]
 \\
 & \hspace{38mm} \times \int_y^{\infty} dy_{n+1}\, \sigma(x_n;\rho)\,e^{-\sigma(x_n;\rho)(y_{n+1}-y_n)}\,\delta(\xt-\xt_n)\Bigg\}.
\end{aligned}  
\label{eq:DIterSol2}
\end{equation}

The main obstacle to applying a random walk algorithm directly to the above equation is the integrated kernel $\sigma(\xt_{i-1};\rho)$,
which cannot be calculated analytically. The analytical form given on RHS of Eq.~(\ref{eq:integraged_kernel}) is obtained with
the following simplification of the integration domain:
\begin{equation}
  \label{eq:approxsig}
\theta\left((\rt' - \rt_i)^2 - \rho^2\right)\:\rightarrow\: \theta(r_i^2 - {r'}^2 - \rho^2 ),
\end{equation}
which means that the UV boundary does not depend on the azimuthal angle between the two dipole vectors. As a consequence,
the length of the emitted dipole $\rt'$ is smaller than the length of the emitter dipole $\rt_i$: $r' < r_i$, 
although the length of the emitter dipole after such an emission can be greater than before the emission, 
i.e.\ $\exists_{\rt'}: |\rt_i - \rt'| > r_i$.
However, the evolution with the approximate integrated kernel given by the RHS of Eq.~(\ref{eq:integraged_kernel}) 
does not include the cases of dipole splitting where ${r'}^2 > r_i^2 - \rho^2$ 
and $\rt_i - \rt'$ are still in the allowed domain, i.e.\ it satisfies the condition $\rho^2 < (\rt_i - \rt')^2 < \Delta_{\rm IR}^2$,
where $\Delta_{\rm IR}$ is some infra-red (IR) cut-off $\sim 1/\Lambda_{\rm QCD}$.  
  
 In order to take into account the above missing dipole splittings, let us first rewrite Eq.~(\ref{eq:integraged_kernel}) 
 in our notation used for a MCMC-method solution:
 \begin{equation}
  \label{eq:KerIntEx}
  \begin{aligned}
    \sigma(x_{i-1};\rho)  & = \int d^2\, \rt_i \,\bar{K}(\xt_{i-1}-\rt_i, \rt_i)\,\theta(r_i^2 - \rho^2 )\,
    \theta(\Delta_{\rm IR}^2 - r_i^2)\,\\
    & \hspace{15mm}
    \times \theta\left((\xt_{i-1}-\rt_i)^2 - \rho^2\right) 
    \theta(\Delta_{\rm IR}^2 - (\xt_{i-1}-\rt_i)^2)\, 
   \\
    & = \sigma_1(x_{i-1};\rho) + \sigma_2(x_{i-1};\rho),
  \end{aligned}
\end{equation}
where
\begin{align}
\label{eq:Kbar}
\bar{K}(\xt_{i-1}-\rt_i, \rt_i) \equiv\; & \frac{\asb}{2\pi}\frac{x_{i-1}^2}{(\xt_{i-1} - \rt_i)^2\,r_i^2} 
= \frac{\asb}{2\pi}\frac{x_{i-1}^2}{x_i^2\,r_i^2}\,, \\
\label{eq:sigma1}
\sigma_1(x_{i-1};\rho) \equiv & \int d^2 \rt_i \,\bar{K}(\xt_{i-1}-\rt_i, \rt_i)\,\theta(r_i^2 - \rho^2 )\,\theta(x_{i-1}^2 - r_i^2 - \rho^2) \nonumber \\
= & \:\asb\ln\frac{ x_{i-1}^2 - \rho^2}{\rho^2}\,,\\
\label{eq:sigma2}
\sigma_2(x_{i-1};\rho) \equiv & \int d^2 \rt_i\, \bar{K}(\xt_{i-1}-\rt_i, \rt_i) \,\theta(r_i^2 - \rho^2 )  \,
\nonumber \\
& \times \bigg[ \theta(\Delta_{\rm IR}^2 - r_i^2)\,\theta\left((\xt_{i-1} - \rt_i)^2 - \rho^2\right)\, 
  \theta(\Delta_{\rm IR}^2 - (\xt_{i-1}-\rt_i)^2)
\nonumber \\
&  \hspace{6mm} 
- \theta(x_{i-1}^2 - r_i^2 - \rho^2)\bigg].
\end{align}

Unfortunately, the integral in Eq.~(\ref{eq:sigma2}) cannot be calculated analytically, thus we need to find 
some simplified kernel $\tilde{K}_2(\xt_{i-1}-\rt_i, \rt_i)$ and a simpler integration domain, such that it completely contains
the integration domain of  Eq.~(\ref{eq:sigma2}), and then apply a Monte Carlo weight that compensates for these simplifications. 
One of the possible choices is the following:
\begin{align}
\label{eq:Ktilde}
\tilde{K}_2(\xt_{i-1}-\rt_i, \rt_i;\rho) =\; & \frac{\asb}{2\pi}\frac{x_{i-1}^2}{\rho^2\,r_i^2} \,,\\
\label{eq:sig2til}
\tilde{\sigma}_2(x_{i-1};\rho) = & \int d^2 \rt_i\, \tilde{K}_2(\xt_{i-1}-\rt_i, \rt_i;\rho) \,\theta(r_i^2 - x_{i-1}^2 + \rho^2 )  
\,\theta(\Delta_{\rm IR}^2 - r_i^2)
\nonumber \\
 = &\: \asb \int dr_i^2 \, \frac{x_{i-1}^2}{2\rho^2\,r_i^2} \int_0^{2\pi} \frac{d\phi_i}{2\pi} 
 =  \asb \frac{x_{i-1}^2}{2\rho^2}\ln\frac{\Delta_{\rm IR}^2}{x_{i-1}^2 - \rho^2}\,.
\end{align}

Then, in Eq.~(\ref{eq:DIterSol2}) we do the following replacements: 
\begin{equation}
\label{eq:sigsimp}
\sigma(x_{i-1};\rho) \rightarrow \tilde{\sigma}(x_{i-1};\rho) = \sigma_1(x_{i-1};\rho) + \tilde{\sigma}_2(x_{i-1};\rho),
\end{equation}
\begin{equation}
\label{eq:Ksimp}
K(\xt_i,\rt_i;\rho) \rightarrow \tilde{K}(\xt_i,\rt_i;\rho) =
\left\{ 
\begin{tabular}{ll}
$\bar{K}(\xt_i,\rt_i),$ & $\rho^2  < r_i^2 < x_{i-1}^2 - \rho^2,$ \\  
 & \\
$\tilde{K}_2(\xt_i,\rt_i;\rho), $ & $x_{i-1}^2 - \rho^2 < r_i^2 < \Delta_{\rm IR}^2.$\\
\end{tabular}
\right.
\end{equation}

Let us start a description of our MCMC algorithm 
by defining probability distribution functions for the main dipole variables, i.e.\ the rapidity $y_i$ and the 2-vector $\rt_i$. 
The probability distribution function ({\it p.d.f.}) for $y_i$ is
\begin{equation}
  \label{eq:yipdf}
\tilde{\eta}(y_i) =  \tilde{\sigma}(x_{i-1};\rho)\,e^{-\tilde{\sigma}(x_{i-1};\rho)(y_i-y_{i-1})},\qquad y_i\in [y_{i-1},+\infty),
\end{equation}
while for $\rt_i$ we have
\begin{equation}
  \label{eq:ripdf}
\tilde{\zeta}(\rt_i) =  \frac{\tilde{K}(\xt_i,\rt_i;\rho)}{\tilde{\sigma}(x_{i-1};\rho)}, \qquad  \rho^2 < r_i^2 < \Delta_{\rm IR}^2.
\end{equation}

Now we can write the approximate iterative solution of Eq.~(\ref{eq:DIterSol2}) using the above {\it p.d.f.}s:
\begin{equation}
\begin{aligned}
 \tilde{D(}\xt,y)  =\: & \int d^2\xt_0\, D(\xt_0,y_0) \Bigg\{ \int_y^{\infty} dy_1 \,\tilde{\eta}(y_1)\,\delta(\xt-\xt_0) 
 \\
 & \hspace{33mm}  + \sum_{n=1}^{\infty} 
  \prod_{i=1}^n \left[ \int_{y_{i-1}}^y dy_i \,\tilde{\eta}(y_i) \int d^2\rt_i  \,\tilde{\zeta(}\rt_i)\right]
 \\
 & \hspace{45mm} \times \int_y^{\infty} dy_{n+1}\, \tilde{\eta}(y_{n+1})\,\delta(\xt-\xt_n)\Bigg\}.
\end{aligned}  
\label{eq:DIterSolpdf}
\end{equation}

Rapidity $y_i$ can easily be generated from the cumulative distribution function
\begin{equation}
  \label{eq:yicdf}
\Xi(y_i) =  \int_{y_{i-1}}^{y_i} dy_i'\,\tilde{\eta}(y_i') = 
1 - e^{-\tilde{\sigma}(x_{i-1};\rho)(y_i-y_{i-1})},
\end{equation}
according to the formula
\begin{equation}
  \label{eq:yigen}
y_i =  y_{i-1} - \frac{\ln R_1}{\tilde{\sigma}(x_{i-1};\rho)}, \qquad R_1\in {\cal U}(0,1).
\end{equation}

Generating $\rt_i$ is more involved. It is a two-dimensional random variable and its {\it p.d.f.} has two branches according
to Eq.~(\ref{eq:Ksimp}). Moreover, the second branch includes an approximation which needs to be compensated by an appropriate
Monte Carlo weight.

We start by selecting a Monte Carlo branch, i.e.\ we generate the random number $R_2\in {\cal U}(0,1)$, and if
\begin{equation}
  \label{eq:ibranch}
 R_2\, \tilde{\sigma}(x_{i-1};\rho) < \sigma_1(x_{i-1};\rho), 
\end{equation}
we select branch {\bf 1}, otherwise we select branch {\bf 2}.

For branch {\bf 1} we define
\begin{equation}
  \label{eq:rifipdf}
\tilde{\zeta}(\rt_i) d^2\rt_i =  f_1(\phi_i | r_i^2)d\phi_i\,g_1(r_i^2)dr_i^2,
\end{equation}
where 
\begin{align}
 g_1(r_i^2) & = \int_0^{2\pi}d\phi_i\,\tilde{\zeta}(\rt_i) 
= \frac{1}{\sigma(x_{i-1};\rho)} \frac{\asb}{2\pi} \int_0^{2\pi}d\phi_i\,\frac{x_{i-1}^2}{2(x_{i-1}^2-2x_{i-1}r_i\cos\phi_i+r_i^2)r_i^2}
\nonumber\\
 \label{eq:gri2}
 & = \frac{1}{2}\left[\ln\frac{x_{n-1}^2-\rho^2}{\rho^2}\right]^{-1}\left[\frac{1}{r_i^2} + \frac{1}{x_{i-1}^2-r_i^2}\right],
\end{align}
and its cumulative distribution function ({\it c.d.f.})
\begin{equation}
  \label{eq:ri2cdf}
  \begin{aligned}
G_1(r_i^2) & =  \int_{\rho^2}^{r_i^2} ds_i^{2}\, g_1(s_i^{2}) = 
\frac{1}{2}\left[\ln\frac{x_{n-1}^2-\rho^2}{\rho^2}\right]^{-1}\left[\ln\frac{r_i^2}{\rho^2} + \ln\frac{x_{i-1}^2 - \rho^2}{x_{i-1}^2 - r_i^2}\right]
\\& = \frac{1}{2}\left[\tilde{G}_1(r_i^2) + \tilde{G}_1(x_{i-1}^2 - r_i^2)\right].
\end{aligned}
\end{equation}
From this distribution, we can generate $r_i^2$ according to
\begin{equation}
  \label{eq:ri2gen}
  r_i^2 =  \left\{ 
  \begin{tabular}{ll}
  $\bar{r}_i^2 \equiv \rho^2\left[\frac{x_{i-1}^2-\rho^2}{\rho^2}\right]^{R_3} $, & $\mathrm{if} \; R_4 \leq 0.5,$\\
  & \\
  $x_{i-1}^2 -  \bar{r}_i^2 ,$ & $\mathrm{if} \; R_4 > 0.5,$
 \end{tabular}
 \right.
\end{equation}
where $R_3,R_4\in {\cal U}(0,1)$.
The conditional {\it p.d.f.} $f_1(\phi|r_i^2)$ can be expressed as a sum of two distributions
\begin{equation}
  \label{eq:fphipdf}
f_1(\phi_i | r_i^2) = h_1(\phi_i | r_i^2) + h_1(2\pi - \phi_i | r_i^2),
\end{equation}
where
\begin{equation}
  \label{eq:hphipdf}
h_1(\phi_i | r_i^2) = \frac{x_{i-1}^2 -r_i^2}{\pi(x_{i-1}^2 -2x_{i-1}r_i\cos\phi_i + r_i^2)},\qquad \phi_i\in [0,\pi).
\end{equation}
Its {\it c.d.f.} reads
\begin{equation}
  \label{eq:Hphipdf}
H_1(\phi | r_i^2) = \int_0^{\phi_i} d\phi_i' \,h(\phi_i' | r_i^2) = \frac{2}{\pi} \arctan\left(\frac{x_{i-1} + r_i}{x_{i-1} - r_i}\tan\frac{\phi_i}{2}\right).
\end{equation}
Using the standard Monte Carlo method of the {\it c.d.f.} inversion, we obtain a formula to generate the random
variable $\phi_i$:
\begin{equation}
  \label{eq:phigen}
  \phi_i =  \left\{ 
  \begin{tabular}{ll}
  $\bar{\phi}_i \equiv 2\arctan\left(\frac{x_{i-1} - r_i}{x_{i-1} + r_i}\tan\frac{\pi R_5}{2}\right)$, & $\mathrm{if} \; R_6 \leq 0.5,$\\
  & \\
  $2\pi - \bar{\phi}_i$, & $\mathrm{if} \; R_6 > 0.5,$
 \end{tabular}
 \right.
\end{equation}
where $R_5,R_6\in {\cal U}(0,1)$.

Having generated $r_i^2$ ($r_i = \sqrt{r_i^2}$) and $\phi_i$, we can construct the 2-vector $\rt_i$: 
\begin{equation}
  \label{eq:rixy}
\left\{ 
 \begin{tabular}{l}
$\left(\rt_i\right)_x  = \frac{r_i}{x_{i-1}}\left[\left(\xt_{i-1}\right)_x \cos\phi_i - \left(\xt_{i-1}\right)_y \sin\phi_i \right], $\\
\\
$\left(\rt_i\right)_y  = \frac{r_i}{x_{i-1}}\left[\left(\xt_{i-1}\right)_x \sin\phi_i + \left(\xt_{i-1}\right)_y \cos\phi_i \right], $
 \end{tabular}
 \right.
\end{equation}
and then the 2-vector $\xt_i$:
\begin{equation}
 \label{eq:xixy}
\left\{ 
 \begin{tabular}{l}
$\left(\xt_i\right)_x  = \left(\xt_{i-1}\right)_x - \left(\rt_i\right)_x,$\\
\\
$\left(\xt_i\right)_y  = \left(\xt_{i-1}\right)_y - \left(\rt_i\right)_y.$
\end{tabular}
 \right.
 \end{equation}

Since in this branch, both $r_i^2$ and $\phi_i$ are generated without any approximation, the compensating 
Monte Carlo weight in this case is
\begin{equation}
  \label{eq:wgt1}
w_i(x_i;\rho) = 1.
\end{equation}

For branch {\bf 2}, we obtain the following {\it p.d.f.}s for the variables $r_i^2$ and $\phi_i$:
\begin{align}
\label{eq:gtilde}
g_2(r_i^2) = \:& \left[\ln\frac{\Delta_{\rm IR}^2}{x_{i-1}^2 - \rho^2}\right]^{-1} \frac{1}{r_i^2}\,, \qquad 
x_{i-1}^2 - \rho^2 \leq r_i^2 <   \Delta_{\rm IR}^2,\\
\label{eq:htilde}
h_2(\phi_i) =\: & \frac{1}{2\pi}\,, \hspace{35mm} 0 \leq \phi_i < 2\pi\,,
\end{align}
from which we can generate the random variables $r_i^2$ and $\phi_i$ as follows:
\begin{align}
\label{eq:ri2gen2}
r_i^2 = \:& (x_{i-1}^2 - \rho^2) \left(\frac{\Delta_{\rm IR}^2}{x_{i-1}^2 - \rho^2}\right)^{R_7}, \qquad 
R_7 \in {\cal U}(0,1),\\
\label{eq:phigen2}
\phi_i  =\: &  2\pi R_8\,, \qquad R_8 \in {\cal U}(0,1).
\end{align}

Having generated $r_i^2$ and $\phi_i$, we can construct the 2-vectors $\rt_i$ and $\xt_i$ according to Eqs.~(\ref {eq:rixy})
and (\ref {eq:xixy}). Then we can calculate the compensating weight
\begin{equation}
\label{eq:wgt2}
 w_i(x_i;\rho) =  \left\{ 
  \begin{tabular}{ll}
  $\frac{\bar{K}(\xt_{i-1}-\rt_i, \rt_i)}{\tilde{K}_2(\xt_{i-1}-\rt_i, \rt_i;\rho)} = \frac{\rho^2}{x_i^2} \: < 1$, & 
  for~ $\rho^2 < x_i^2 < \Delta_{\rm IR}^2$, \\
  & \\
   $0$, & otherwise.\\
 \end{tabular}
 \right.
\end{equation}
 The weight $w_i(x_i;\rho)$ is bounded from above by $1$ which is a good feature that allows the application of a veto algorithm.

Now we can propose the following MCMC algorithm for numerical evaluation of  Eq.~(\ref{eq:DIterSolpdf}):
\begin{itemize}
\item[{\bf Step 1:}]
Set $i=0$. Choose the starting rapidity value $y_0$ and generate the initial dipole vector $\xt_0$ according to 
the distribution $D(\xt_0,y_0)$.
Typical choices are $y_0=0$ and $D(\xt_0,y_0)=\delta(\xt_0-\rt_1)$, for some fixed value of $\rt_1$. 
Then go to step~{\bf 2}.
\item[{\bf Step 2:}]
Set $i:=i+1$ and generate the random variable $y_i$ according to Eq.~(\ref{eq:yigen}). 
If $y_i > y$, then {\bf stop}, otherwise go to step~{\bf 3}.
\item[{\bf Step 3:}]
Select the $\rt_i$ branch according to Eq.~(\ref{eq:ibranch}). For branch {\bf 1}, 
generate random variables $r_i^2$ and $\phi_i$ according to Eqs.~(\ref{eq:ri2gen}) and (\ref{eq:phigen}),
while for branch {\bf 2} according to Eqs.~(\ref{eq:ri2gen2}) and (\ref{eq:phigen2}), respectively.
Construct the 2-vectors $\rt_i$ and $\xt_i$ according to Eqs.~(\ref{eq:rixy}) and (\ref{eq:xixy}). 
Calculate the Monte Carlo weight according to Eq.~(\ref{eq:wgt1}) or Eq.~(\ref{eq:wgt2}), depending on the selected branch.
Apply the veto algorithm, i.e.\ generate a random number $R_9\in {\cal U}(0,1)$, and if
\begin{equation}
  \label{eq:VetoAlg}
 R_9 \leq w_i(x_i;\rho),
\end{equation}
accept the computed variables $(\rt_i,\xt_i)$, otherwise set: $\rt_i=\mathbf{0},\, \xt_i=\xt_{i-1}$.
Then go to step~{\bf 2}.
\end{itemize}
Repeat the above algorithm a desired number of times $N$, histograming the obtained values of $\xt$ for each Monte Carlo event. 
In the end normalise the histogram, i.e.\ divide each bin contents by $N$ times a bin size. 
The resulting histogram will correspond to the distribution $D(\xt,y)$ of Eq.~(\ref{eq:DIterSolpdf}) 
which is the solution of the differential equation (\ref{eq:dDxdy}).
 
The above MCMC algorithm has been implemented in a C++ program called {\sf DIPMAR}~\cite{DIPMAR}.

\subsubsection{Number of dipoles}
\label{sssec:ndip}

 The presented MCMC algorithm corresponds to the solution of the evolution equation for the probability
distribution $D(\xt,y)$ of the dipole $\xt$ with rapidity $y$ from some initial distribution $D(\xt_0,y_0)$
 of the dipole $\xt_0$ at rapidity $y_0$. In this evolution, only a single branch in which the first dipole subsequently splits 
 emitting new dipoles is considered, while splittings of its off-spring dipoles are not generated -- these splittings are
 integrated out, as they do not contribute to the distribution $D(\xt,y)$. 
 This is similar to the evolution of parton distribution functions in a nucleon e.g.\ via the DGLAP equations.
 
 If we are interested in the evolution of the number of dipoles, we need to take into account also splittings of all off-spring dipoles.
 We can generalize the dipole distribution function of the dipole $\rt_1$ given in Eq.~(\ref{eq:Dxy}) to any dipole $\rt_k$:
\begin{equation}
  \label{eq:Dxky}
  D(\xt^{(k)},y;\rho)  =  \sum_{n=1}^\infty \prod_{i=1}^n\int d^2 \rt_i   \delta(\xt^{(k)}-\rt_k) P_{n}(y, \rt_1, \ldots, \rt_n;\rho),\quad
  k = 1,2,\ldots.
 \end{equation}
Since
\begin{equation}
 \label{eq:DkdyNorm}
\int d^2\xt^{(k)} \, D(\xt^{(k)},y;\rho)  = 1,  
\end{equation}
the splittings of off-spring dipoles do not affect the evolution of the dipole distribution function of a parent dipole.
Therefore, if we are interested only in the dipole distribution function of the original dipole $\rt_1$, we
do not need to perform splitting of any of its off-springs, and use the MCMC algorithm described in the previous subsections.
However, if our aim is to evaluate the number of dipoles at rapidity $y$, then we have to generate dipole emissions also 
from all the off-spring dipoles. 

Let us generalize the iterative solution given in Eq.~(\ref{eq:DIterSol2}) to any dipole $\rt_k$:
\begin{equation}
\begin{aligned}
 D(\xt^{(k)} ,y;\rho)  =\: & \int d^2\xt_0\, D(\xt_0^{(k)} ,y_0^{(k)};\rho) \Bigg\{ \int_y^{\infty} dy_1^{(k)}  \,\sigma(x_0^{(k)} ;\rho)
 \,e^{-\sigma(x_0^{(k)} ;\rho)(y_1^{(k)} -y_0^{(k)})}\,\delta(\xt^{(k)} -\xt_0^{(k)}) 
\\
  &\hspace{16mm}  
  + \sum_{n=1}^{\infty} \prod_{i=1}^n \left[ \int_{y_{i-1}^{(k)} }^y dy_i^{(k)}  \int d^2\rt_i^{(k)}  \,K(\xt_i^{(k)} ,\rt_i^{(k)}; \rho) \,
  e^{-\sigma(\xt_{i-1}^{(k)} ;\rho)(y_i^{(k)} -y_{i-1})^{(k)} }\right]
 \\
 & \hspace{26mm} \times \int_y^{\infty} dy_{n+1}^{(k)} \, \sigma(x_n^{(k)} ;\rho)\,
 e^{-\sigma(x_n^{(k)} ;\rho)(y_{n+1}^{(k)} -y_n^{(k)})}\,\delta(\xt^{(k)} -\xt_n^{(k)} )\Bigg\},
\end{aligned}  
\label{eq:DkIterSol2}
\end{equation}
where $\xt_0^{(k)}$ and $y_0^{(k)}$ are the initial values of the dipole vector and the rapidity at the moment of creation
of a given dipole: for the first dipole they are the starting values $\xt_0$ and $y_0$, while for any off-spring dipoles
they correspond to the dipole vector and rapidity values of the moment of emission of that off-spring dipole.   

To calculate the number of dipoles, we need to perform the evolution of each dipole from the 
starting values ($\xt_0^{(k)},y_0^{(k)}$) at its creation moment to the final rapidity value $y$, common for all dipoles. 
This means that the MCMC algorithm described above should be
repeated for every created dipole until no more emissions occur.  There are two main ways to do this: 
\begin{enumerate}
\item
An algorithm based on a binary tree can be applied in which a node corresponds 
to a dipole that splits into two off-spring dipoles. In this case, one starts the MCMC evolution from a single node at
the top of the tree corresponding to the initial dipole and goes down the binary tree generating dipole splitting at
subsequent tree levels. If no more splitting is allowed at a given node, i.e.\ $y_{n+1}^{(k)} \ge y$, then this node becomes
a leaf of a three. The algorithm continues until all nodes of the tree become leafs, and the number of
dipoles is equal to the number of leafs in this tree.
\item
The other option is an algorithm based on a genealogical tree, in which one starts from the initial dipole as the top node
(root) of the tree, then all its off-springs are generated, i.e.\ the MCMC evolution of the initial dipole is performed until 
$y_{n+1}^{(1)} \ge y$, and then the same is repeated for each of its off-springs following the order of their creation and stopping
each evolution if $y_{n+1}^{(k)} \ge y$, and similarly for their off-springs, etc. The algorithm ends when there are no more
dipoles to evolve, i.e.\ all of them reached the final rapidity value $y$. Here, the subsequent levels of the tree
correspond to generations of the genealogical tree, i.e.\ at the top is the main parent, in the next level all its children,
then its grandchildren, and so on. The number of dipoles in this case is equal to the number of nodes in such a tree.
\end{enumerate}

Our Monte Carlo generator {\sf DIPMAR}~\cite{DIPMAR} implements the second algorithm because it is easier to combine it 
with the MCMC evolution algorithm described above. We use the latter algorithm in a loop,
starting from the evolution of the initial dipole and adding to a data base, in the form of a singly linked list, 
the information on the values of ($\xt_0^{(k)},y_0^{(k)}$) for each created off-spring dipole, 
then performing the evolution of the next element in the list, again adding to
the list the corresponding information on its off-springs, until the end of the list is reached.  
The number of dipoles $n$ is equal to the number of elements in the list, including the initial dipole.

The probability distribution of the number of dipoles $p_{n}(y;\rho), n = 1,2,\ldots,$ can be obtained by histograming 
the number of generated dipoles $n = 1,2,\ldots,n_{\rm max}$, 
where $n_{\rm max}$ should be chosen such that $\sum_{n=n_{\rm max}}^{\infty} p_{n}(y;\rho) \ll 1$,
and then normalizing such histograms to $1$, i.e.\ dividing each bin contents by the number of generated events $N$.

\subsubsection{DLLA}
\label{sssec:dlla}

The double leading-log approximation (DLLA) can be included in the MCMC algorithm described above
after applying the following modifications:
 \begin{align}
  \label{eq:MMCKdlla}
\bar{K}(\xt_i,\rt_i) & \rightarrow 
\bar{K}_{\rm DLLA}(\xt_i,\rt_i) =  \frac{\asb}{2\pi}\frac{1}{r_i^2},\quad \rho^2 < r_i^2 < x_{i-1}^2,\\
  \label{eq:MMCsigdlla}
\sigma(x_{i-1};\rho) & \rightarrow \sigma_{\rm DLLA}(x_{i-1};\rho) =  \asb \ln\frac{x_{i-1}^2}{\rho^2}, \\
  \label{eq:MMCphidlla}
\phi_i & = 0,\\
  \label{eq:MMCxidlla}
\xt_i & = \xt_{i-1}.
\end{align}
This corresponds to one-dimensional dipole evolution in which the emitter dipole size remains unchanged after the emission.
In this case, the random variables $y_i$ and $\rt_i$ can be easily generated without any approximation.

Such an option has been implemented in the Monte Carlo generator {\sf DIPMAR}~\cite{DIPMAR}.
We have checked that for the mean dipole multiplicity $\langle n \rangle$, it agrees very well with the analytical solution
of Eq.~(\ref{eq:BesselI0}); see Appendix~B.3.

\subsection{Monte Carlo program {\sf LLMC}}
%\subsection{Direct Monte-Carlo algorithm for the Levin--Lublinsky equations}
\label{App:B2}
The Levin--Lublinsky equations given in Eq.~(\ref{eq:dpn}) can be solved directly via a Monte Carlo algorithm. 
To this end, the set of equations can be written as
\begin{align}
  e^{-Y \Delta_n(\vec{r}_1,\ldots,\vec{r}_n)}\frac{d}{dY}P_n(Y ; \vec{r}_1,\ldots,\vec{r}_n)e^{Y \Delta_n(\vec{r}_1,\ldots,\vec{r}_n)}
  =\sum_{i=1}^{n-1} K(\vec{r}_i, \vec{r}_n)\, P_{n-1}(Y; \vec{r}_1,\ldots,\vec{r}_i+\vec{r}_n,\ldots\vec{r}_{n-1}),
  \label{eq:dpnSud}
  \end{align}
  with
\begin{equation}
\Delta_n(\vec{r}_1,\ldots,\vec{r}_n)\equiv \sum_{i=1}^n\sigma_{\rm IR}(\vec{r}_i)\,,
\end{equation}
where 
\begin{equation}
\sigma_{\rm IR}(\vec{r}_i)=\int_{\mathbb{R}^2}d^2r'K(\vec{r}', \vec{r_i}-\vec{r}')\theta(r_{\rm max}^2-{r'}^2) \theta(r_{\rm max}^2-(\vec{r_i}-\vec{r}')^2)\,,
\end{equation}
is the infrared-regulated integral over the kernel $K$, with the IR-regulator $r_{\rm max}\sim 1/\Lambda_{\rm QCD}$.
Integration of Eq.~(\ref{eq:dpnSud}) over a rapidity interval $Y\in [Y_A,Y_B]$  yields
\begin{align}
  P_n(Y_B ; \vec{r}_1,\ldots,\vec{r}_n)
  &=e^{-(Y_B-Y_A) \Delta_n(\vec{r}_1,\ldots,\vec{r}_n)}P_n(Y_A ; \vec{r}_1,\ldots,\vec{r}_n)e^{y \Delta_n(\vec{r}_1,\ldots,\vec{r}_n)}\nonumber\\
  &+\int_{Y_A}^{Y_B}dY e^{-(Y_B-Y) \Delta_n(\vec{r}_1,\ldots,\vec{r}_n)} \sum_{i=1}^{n-1} K(\vec{r}_i, \vec{r}_n)\, P_{n-1}(Y; \vec{r}_1,\ldots,\vec{r}_i+\vec{r}_n,\ldots\vec{r}_{n-1}).
  \label{eq:dpnSudInt}
\end{align}
With the above equation, a Monte Carlo algorithm that samples the probability distributions $P_n$ for different rapidity values can be inferred: In order to sample $P_n$ at a given rapidity, a set of dipoles
$\{\vec{r}_1,\ldots,\vec{r}_n\}$ is selected. The algorithm starts with a given sample for $P_1$, the dipole $\vec{r}_1$ for rapidity $Y_0=0$, and then obtains a rapidity $Y_1$ for splitting into two dipoles $\vec{r}_2$ and $\vec{r}_3$, which represents a sample for $P_2$. This procedure is repeated for samples of $P_3$, $P_4$, etc. until a number of dipoles $n_{\rm max}$ is reached at the rapidity $y_{n_{\rm max}}$ above a previously set maximum rapidity $y_{\rm max}$, i.e. once $y_{n_{\rm max}}>y_{\rm max}$ the algorithm stops. Then, the algorithm is repeated for many different events.

The sampling of the distribution $P_{n}(Y_{n-1} ; \vec{r}_1,\ldots,\vec{r}_n)$ from the distribution $P_{n-1}(Y_{n-2} ; \vec{r}_1,\ldots,\vec{r}_{n-1})$ can be done in the following way:
\begin{enumerate}
\item First, the rapidity $Y_{n-1}$ is selected by choosing randomly a real number $R_y\in {\cal U}(0,1)$, i.e.\ from the uniform distribution over the range $(0,1)$.
Then the equation 
\begin{equation}
R_y=e^{-(Y_{n-1}-Y_{n-2})\Delta_{n-1}(\vec{r}_1,\ldots,\vec{r}_{n-1})}
\end{equation}
is solved for $Y_{n-1}$.
\item Then, one needs to determine which dipole branches. If there are $n-1$ dipoles, any dipole $i$ can split with the probability 
\begin{equation}
\frac{\sigma_{\rm IR}(\vec{r}_i)}{\Delta_{n-1}(\vec{r}_1,\ldots,\vec{r}_{n-1})}.
\end{equation}
In the Monte Carlo algorithm, the corresponding random selection is made.
\item A dipole $i$ splits into two dipoles $\tilde{i}$ and $n$. The length ${r}_{\tilde{i}}$ of the dipole $\tilde{i}$ is selected 
by solving the equation
\begin{equation}
 R_r=\frac{\int_\rho^{{r}_{\tilde{i}}}r'dr'\int_0^{2\pi}d\phi K(\vec{r}',\vec{r}_i- \vec{r}')}{\sigma_{\rm IR}(\vec{r}_i)}
 \label{eq:llmcrseleq}
\end{equation}
for ${r}_{\tilde{i}}$, where the number $R_r\in {\cal U}(0,1)$, and the azimuthal angle $\phi$ is defined by 
\begin{equation}
r_ir'\cos\phi=\vec{r}_i\cdot\vec{r}'\,.
\end{equation} 
\item The azimuthal angle ${\phi}_{\tilde{i}}$ of the dipole $\tilde{i}$ is selected 
by solving the equation
%\begin{equation}
%R_{\phi}=\frac{\int_0^{\phi_{\tilde{i}}}d\phi\frac{1}{r_i^2+r_{\tilde{i}}^2-2r_ir_{\tilde{i}}\cos(\phi)} \theta(r_i^2+r_{\tilde{i}}^2-2r_ir_{\tilde{i}}\cos(\phi) -\rho^2) \theta(r_{\rm max}^2-r_i^2-r_{\tilde{i}}^2+2r_ir_{\tilde{i}}\cos(\phi) )     }{\int_0^{2\pi}d\phi\frac{1}{r_i^2+r_{\tilde{i}}^2-2r_ir_{\tilde{i}}\cos(\phi)} \theta(r_i^2+r_{\tilde{i}}^2-2r_ir_{\tilde{i}}\cos(\phi) -\rho^2) \theta(r_{\rm max}^2-r_i^2-r_{\tilde{i}}^2+2r_ir_{\tilde{i}}\cos(\phi) )}\,,
%\end{equation}
\begin{equation}
R_{\phi}=\frac{\int_0^{\phi_{\tilde{i}}}d\phi\,\frac{1}{f(r_i,r_{\tilde{i}},\phi)} \,\theta(f(r_i,r_{\tilde{i}},\phi) -\rho^2)\, \theta(r_{\rm max}^2-f(r_i,r_{\tilde{i}},\phi) )     }{\int_0^{2\pi}d\phi\,\frac{1}{f(r_i,r_{\tilde{i}},\phi)} \,\theta(f(r_i,r_{\tilde{i}},\phi) -\rho^2)\, \theta(r_{\rm max}^2-f(r_i,r_{\tilde{i}},\phi) ) )}\,,
 \label{eq:llmcphiseleq}
\end{equation}
for $\phi_{\tilde{i}}$, where $R_{\phi}\in {\cal U}(0,1)$ is randomly chosen, and
\begin{equation}
f(r_i,r_{\tilde{i}},\phi)=r_i^2+r_{\tilde{i}}^2-2r_ir_{\tilde{i}}\cos\phi\,.
\end{equation}
\item The dipole vector $\vec{r}_n$ is obtained via 
\begin{equation}
\vec{r}_n=\vec{r}_i-\vec{r}_{\tilde{i}}.
\end{equation}
\item The new sample for $P_n$ consists of the rapidity $Y_{n-1}$ and the set of dipole vectors: \\
$$\{\vec{r}_1,\ldots,\vec{r}_{i-1},\vec{r}_{\tilde{i}},\vec{r}_{i+1},\ldots,\vec{r}_{n-1},\vec{r}_n\}.$$
\end{enumerate}

The above Monte Carlo algorithm has been implemented in a C++ program called {\sf LLMC}~\cite{LLMC}.
In this program, the relevant integrals are computed numerically as follows:
\begin{itemize}
    \item The values for the integrals in Eq.~(\ref{eq:llmcrseleq}) and $\sigma_{IR}(\vec{r}_i)$ are calculated numerically prior to the simulation of dipole cascades and then are stored in the form of a grid with the numbers $N_{r_0}=600$ and $N_{r_1}=N_{r_0}=600$ of the equidistant bins in $r_i$ and $r_{\tilde{i}}$, respectively, both in the ranges of $r_i,\,r_{\tilde{i}}\in [\rho,\,r_{\rm max}]$. The integrals of Eq.~(\ref{eq:llmcrseleq}) are calculated as the sum of integrals for the equidistant bins in $r'$ and $\phi$, where the integral of a single bin is approximated by the trapezoidal rule. First, for fixed values of $r'$, the integrals over $\phi$ are obtained as a sum of $N_\phi=1000$ trapeze areas. For subsequent integration over $r'$, the bin width  $\Delta r=(r_{\rm max}-\rho)/(2N_{r_1})$ is used. In steps 1, 2 and 3 of the Monte Carlo algorithm, the integrals are calculated from the grid for the tabulated values of the dipole lengths that are closest to the ones obtained by the algorithm for ${r_1,\,\dots,\, r_n}$ and $r_{\tilde{i}}$. This leads to additional discretization errors.
    \item The integrals of both the numerator and the denominator of Eq.~(\ref{eq:llmcphiseleq}) are calculated for particular values of $r_i$ and $r_{\tilde{i}}$ as the sum of the integrals over the equidistant bins of sizes $\Delta \phi=2\pi/N_{\phi}$, where the integrals over the individual bins are approximated as trapeze areas. 
\end{itemize}

\subsubsection{DLLA}
For the implementation of DLLA in {\sf LLMC}, the following changes are made in the algorithm described previously: 
\begin{enumerate}
    \item The function $\sigma_{\rm IR}$ is replaced by a function $\sigma_{\rm DLLA}$, defined as 
    \begin{equation}
        \sigma_{\rm DLLA}(\vec{r}_i)=\bar{\alpha}_s\ln\left(\frac{r_i^2}{\rho^2}\right)\,.
    \end{equation}
    \item In step 3 of the algorithm, Eq.~(\ref{eq:llmcrseleq}) is replaced by 
    \begin{equation}
        R_r=\frac{\sigma_{\rm DLLA}(\vec{r}_{\tilde{i}})}
        {\sigma_{\rm DLLA}(\vec{r}_i)}\,.
    \end{equation}
    The above equation is solved analytically for ${r}_{\tilde{i}}$ under the assumption that $\rho \leq {r}_{\tilde{i}} \leq {r}_{i} $.
    \item Instead of step 4 of the algorithm, it is assumed that $$\phi_{\tilde{i}}=0\,.$$
    \item Step 5 of the algorithm is also changed. First of all, as follows from the previous remark, the dipole lengths, $\vec{r}_{{i}}$, $\vec{r}_{\tilde{i}}$, and $\vec{r}_n$ are all parallel. Furthermore we set
    \begin{equation}
        r_n=r_i\,.
    \end{equation}
\end{enumerate}

\subsection{Comparisons of Monte Carlo solutions}
\label{App:B3}

We have developed two independent Monte Carlo algorithms solving the Levin--Lublinsky equation, and implemented
them in two different computer programs: {\sf DIPMAR} and {\sf LLMC}, as described in Appendices B1 and B2, respectively.
This allows us to cross-check the two solutions in order to get confidence that they are correct in terms of the 
proposed algorithms, their implementations in the pertinent computer codes, and numerical precision of relevant computations. 

\begin{figure}[!ht]
    %\centering     
    \includegraphics[width=1.0\linewidth]{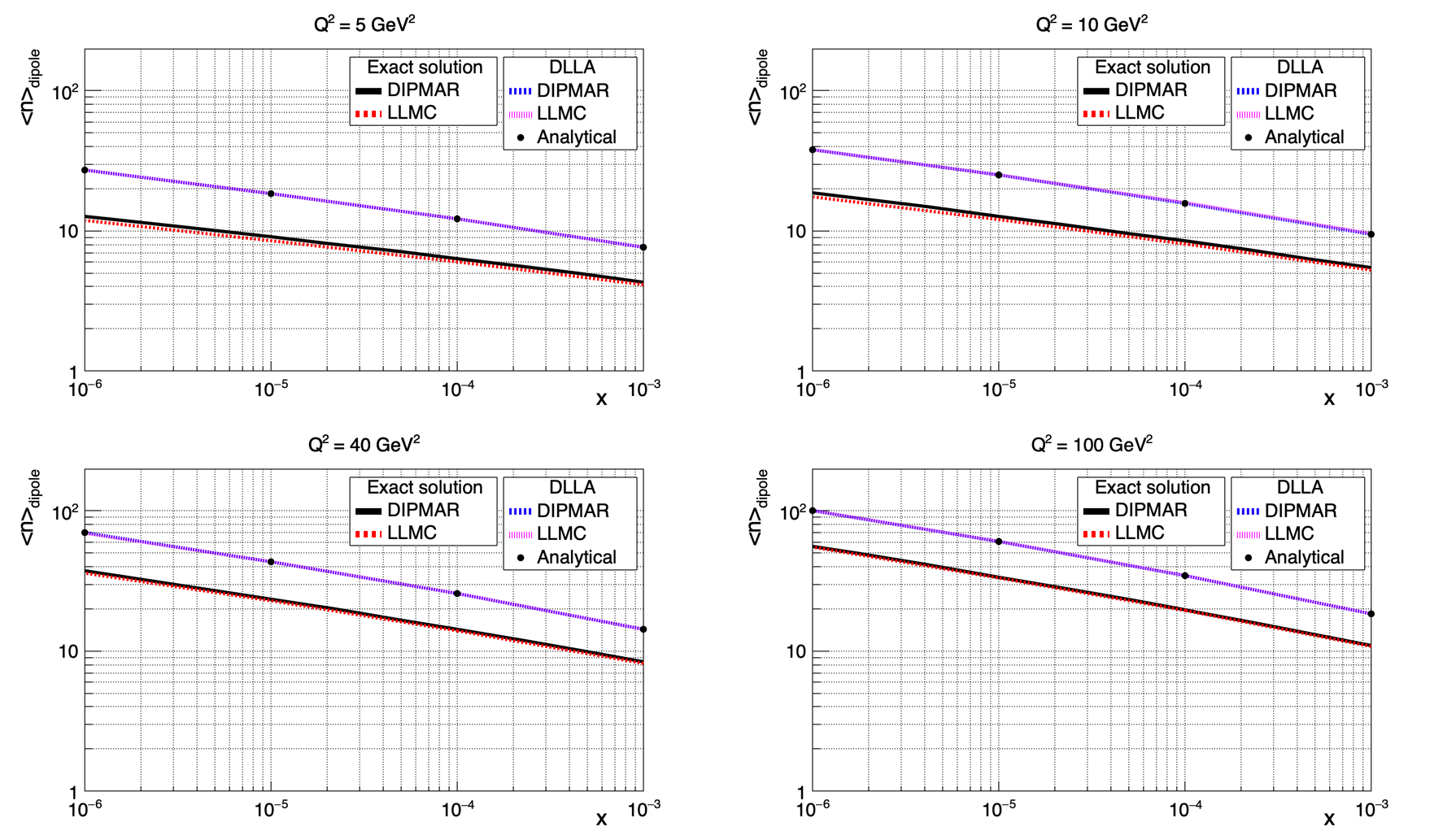}
    \caption{Comparison between the Monte Carlo programs {\sf DIPMAR} and {\sf LLMC} for the mean dipole multiplicity $\langle n \rangle(x,Q^2)$ corresponding to the exact solution of the Levin--Lublinsky equation and its DLLA version,
    for the fixed $\asb = 0.1$ and the initial dipole size $r=0.85\,$fm.
    In the `Exact solution', the IR cut-off $\Delta_{\rm IR} = r = 0.85\,$fm was also applied.
    For DLLA, the results
    of the analytical solution according to Eq.~(\ref{eq:BesselI0}) are also shown (black dots).
     }
        \label{fig:meanmultcomp}
\end{figure}
\begin{figure}[!h]
    %\centering     
    \includegraphics[width=1.0\linewidth]{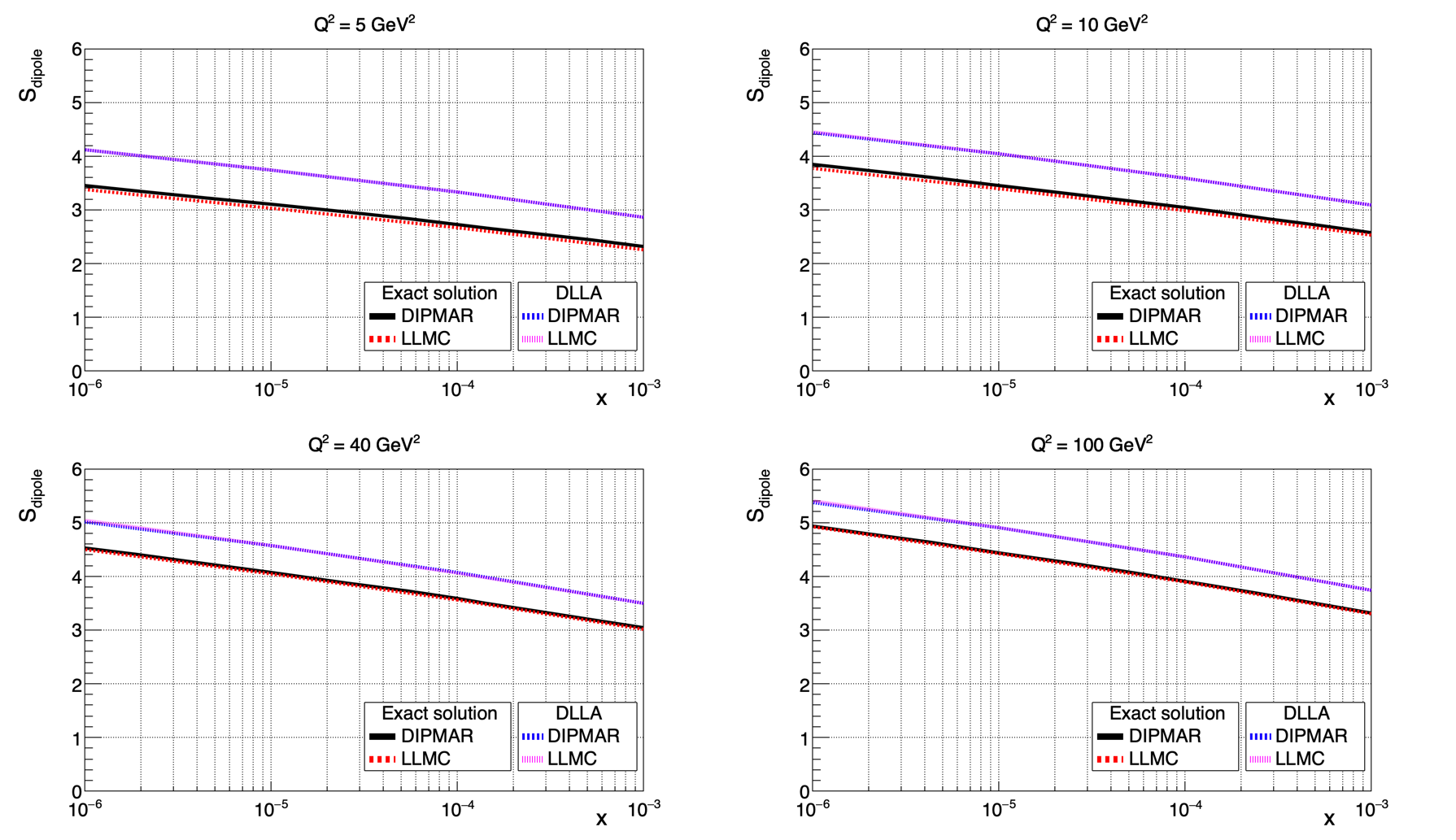}
    \caption{Comparison between the Monte Carlo programs {\sf DIPMAR} and {\sf LLMC} for the dipole entropy $S(x,Q^2)$
    corresponding to the exact solution of the Levin--Lublinsky equation and its DLLA version,
    for the fixed $\asb = 0.1$ and the initial dipole size $r=0.85\,$fm.
    In the `Exact solution', the IR cut-off $\Delta_{\rm IR} = r = 0.85\,$fm was also applied.    
    }
        \label{fig:entropycomp}
\end{figure}

In Figs.~\ref{fig:meanmultcomp} and \ref{fig:entropycomp}, we show comparisons of the two programs for the mean dipole multiplicity $\langle n \rangle(x,Q^2)$ and the dipole entropy $S(x,Q^2)$, respectively. We present the results for both the exact and DLLA solutions,
for the values of $Q^2 = 5, 10, 40, 100\,$GeV$^2$, with the fixed $\asb = 0.1$ and the initial dipole size $r=0.85\,$fm.
In the exact solution, we also applied the IR cut-off $\Delta_{\rm IR} = r = 0.85\,$fm.
As one can see, a good agreement (at the percent level) between the two programs is found for both solutions. 
In addition, in Fig.~\ref{fig:meanmultcomp}, we show the results corresponding to the analytical solution of the DLLA equation for 
the mean dipole multiplicity of Eq.~(\ref{eq:BesselI0})  (black dots) which agree very well with the DLLA solutions of both 
Monte Carlo programs.
These comparisons constitute an important benchmark for our two independent Monte Carlo algorithms used to solve the Levin--Lublinsky
equation, for both the exact and the DLLA cases, as well as their implementation in two different Monte Carlo programs, 
{\sf DIPMAR} and {\sf LLMC}.

\newpage
\bibliographystyle{jhep} 
\bibliography{references}

\end{document}